
\input harvmac
\input amssym



\newfam\frakfam
\font\teneufm=eufm10
\font\seveneufm=eufm7
\font\fiveeufm=eufm5
\textfont\frakfam=\teneufm
\scriptfont\frakfam=\seveneufm
\scriptscriptfont\frakfam=\fiveeufm


\def\bb{
\font\tenmsb=msbm10
\font\sevenmsb=msbm7
\font\fivemsb=msbm5
\textfont1=\tenmsb
\scriptfont1=\sevenmsb
\scriptscriptfont1=\fivemsb
}



\newfam\dsromfam
\font\tendsrom=dsrom10
\textfont\dsromfam=\tendsrom
\def\ds{\fam\dsromfam \tendsrom}


\newfam\mbffam
\font\tenmbf=cmmib10
\font\sevenmbf=cmmib7
\font\fivembf=cmmib5
\textfont\mbffam=\tenmbf
\scriptfont\mbffam=\sevenmbf
\scriptscriptfont\mbffam=\fivembf
\def\mbf{\fam\mbffam \tenmbf}


\newfam\mbfcalfam
\font\tenmbfcal=cmbsy10
\font\sevenmbfcal=cmbsy7
\font\fivembfcal=cmbsy5
\textfont\mbfcalfam=\tenmbfcal
\scriptfont\mbfcalfam=\sevenmbfcal
\scriptscriptfont\mbfcalfam=\fivembfcal


\newfam\mscrfam
\font\tenmscr=rsfs10
\font\sevenmscr=rsfs7
\font\fivemscr=rsfs5
\textfont\mscrfam=\tenmscr
\scriptfont\mscrfam=\sevenmscr
\scriptscriptfont\mscrfam=\fivemscr
\def\scr{\fam\mscrfam \tenmscr}




\def\tilde{\widetilde}
\def\t{\tilde}
\def\hat{\widehat}

\def\bar{\overline}
\def\b{\bar}
\def\bsq#1{{{\b{#1}}^{\lower 2.5pt\hbox{$\scriptstyle 2$}}}}
\def\bexp#1#2{{{\b{#1}}^{\lower 2.5pt\hbox{$\scriptstyle #2$}}}}
\def\dotexp#1#2{{{#1}^{\lower 2.5pt\hbox{$\scriptstyle #2$}}}}


\def\rt2{\sqrt{2}}
\def\half {{1 \over 2}}

\def\d{\partial}

\def\grad{\nabla}

\def\det{\mathop{\rm det}}

\def\vector#1{{\mbf \vec{#1}}}


\font\tenbifull=cmmib10
\font\tenbimed=cmmib7
\font\tenbismall=cmmib5
\textfont9=\tenbifull \scriptfont9=\tenbimed
\scriptscriptfont9=\tenbismall

\mathchardef\bbGamma="7000
\mathchardef\bbDelta="7001
\mathchardef\bbPhi="7002
\mathchardef\bbAlpha="7003
\mathchardef\bbXi="7004
\mathchardef\bbPi="7005
\mathchardef\bbSigma="7006
\mathchardef\bbUpsilon="7007
\mathchardef\bbTheta="7008
\mathchardef\bbPsi="7009
\mathchardef\bbOmega="700A
\mathchardef\bbalpha="710B
\mathchardef\bbbeta="710C
\mathchardef\bbgamma="710D
\mathchardef\bbdelta="710E
\mathchardef\bbepsilon="710F
\mathchardef\bbzeta="7110
\mathchardef\bbeta="7111
\mathchardef\bbtheta="7112
\mathchardef\bbiota="7113
\mathchardef\bbkappa="7114
\mathchardef\bblambda="7115
\mathchardef\bbmu="7116
\mathchardef\bbnu="7117
\mathchardef\bbxi="7118
\mathchardef\bbpi="7119
\mathchardef\bbrho="711A
\mathchardef\bbsigma="711B
\mathchardef\bbtau="711C
\mathchardef\bbupsilon="711D
\mathchardef\bbphi="711E
\mathchardef\bbchi="711F
\mathchardef\bbpsi="7120
\mathchardef\bbomega="7121
\mathchardef\bbvarepsilon="7122
\mathchardef\bbvartheta="7123
\mathchardef\bbvarpi="7124
\mathchardef\bbvarrho="7125
\mathchardef\bbvarsigma="7126
\mathchardef\bbvarphi="7127


\def\alphadot{{\dot\alpha}}


\def\ibar{\b{i}}
\def\jbar{\b{j}}


\def\thetasq{\theta^2}

\def\jbar{\b{j}}


\def\CA{{\cal A}}

\def\CF{{\cal F}}

\def\CK{{\cal K}}
\def\CL{{\cal L}}
\def\CM{{\cal M}}
\def\CN{{\cal N}}

\def\CR{{\cal R}}
\def\CS{{\cal S}}

\def\CU{{\cal U}}


\def\1{{\ds 1}}
\def\R{\hbox{$\bb R$}}
\def\C{\hbox{$\bb C$}}


\def\ep{\varepsilon}

\noblackbox




\lref\IvanovAI{
  S.~Ivanov and G.~Papadopoulos,
  ``Vanishing theorems and string backgrounds,''
Class.\ Quant.\ Grav.\  {\bf 18}, 1089 (2001).
[math/0010038 [math-dg]].
}

\lref\DumitrescuIU{
  T.~T.~Dumitrescu, N.~Seiberg,
  ``Supercurrents and Brane Currents in Diverse Dimensions,''
[arXiv:1106.0031 [hep-th]].
}

\lref\FestucciaWS{
  G.~Festuccia, N.~Seiberg,
  ``Rigid Supersymmetric Theories in Curved Superspace,''
JHEPA,1106,114.\ 2011 {\bf 1106}, 114 (2011).
[arXiv:1105.0689 [hep-th]].
}

\lref\WessCP{
  J.~Wess, J.~Bagger,
  ``Supersymmetry and supergravity,''
Princeton, USA: Univ. Pr. (1992) 259 p.
}

\lref\StelleYE{
  K.~S.~Stelle, P.~C.~West,
  ``Minimal Auxiliary Fields for Supergravity,''
Phys.\ Lett.\  {\bf B74}, 330 (1978).
}

\lref\FerraraEM{
  S.~Ferrara, P.~van Nieuwenhuizen,
  ``The Auxiliary Fields of Supergravity,''
Phys.\ Lett.\  {\bf B74}, 333 (1978).
}

\lref\SohniusTP{
  M.~F.~Sohnius, P.~C.~West,
  ``An Alternative Minimal Off-Shell Version of N=1 Supergravity,''
Phys.\ Lett.\  {\bf B105}, 353 (1981).
}

\lref\SohniusFW{
  M.~Sohnius, P.~C.~West,
  ``The Tensor Calculus And Matter Coupling Of The Alternative Minimal Auxiliary Field Formulation Of N=1 Supergravity,''
Nucl.\ Phys.\  {\bf B198}, 493 (1982).
}

\lref\FerraraPZ{
  S.~Ferrara, B.~Zumino,
  ``Transformation Properties of the Supercurrent,''
Nucl.\ Phys.\  {\bf B87}, 207 (1975).
}

\lref\GatesNR{
  S.~J.~Gates, M.~T.~Grisaru, M.~Rocek, W.~Siegel,
  ``Superspace Or One Thousand and One Lessons in Supersymmetry,''
Front.\ Phys.\  {\bf 58}, 1-548 (1983).
[hep-th/0108200].
}

\lref\KomargodskiRB{
  Z.~Komargodski, N.~Seiberg,
  ``Comments on Supercurrent Multiplets, Supersymmetric Field Theories and Supergravity,''
JHEP {\bf 1007}, 017 (2010).
[arXiv:1002.2228 [hep-th]].
}

\lref\SeibergVC{
  N.~Seiberg,
  ``Naturalness versus supersymmetric nonrenormalization theorems,''
Phys.\ Lett.\  {\bf B318}, 469-475 (1993).
[hep-ph/9309335].
}

\lref\Kosmann{
  Y.~Kosmann,
  ``D\'eriv\'ees de Lie des spineurs,''
Ann.\ di Matematica Pura e Appl.\  {\bf 91}, 317Ð395 (1972).
}

\lref\Pontecorvo{
M.~Pontecorvo,
  ``Complex structures on Riemannian four-manifolds"
Math.\ Ann.\   {\bf 309}, 159-177 (1997).
}

\lref\Boyertwo{
C.~P.~Boyer,
  ``A note on hyper-Hermitian four-manifolds"
Proc.\ Amer.\ Math.\ Soc.\ {\bf 102}, 157-164 (1988).
}

\lref\HamaEA{
  N.~Hama, K.~Hosomichi and S.~Lee,
  ``SUSY Gauge Theories on Squashed Three-Spheres,''
JHEP {\bf 1105}, 014 (2011).
[arXiv:1102.4716 [hep-th]].
}

\lref\WittenZE{
  E.~Witten,
  ``Topological Quantum Field Theory,''
Commun.\ Math.\ Phys.\  {\bf 117}, 353 (1988).
}

\lref\Oldmin{
T.T.~Dumitrescu, G.~Festuccia, and~N.~Seiberg, to appear.
}

\lref\kilspman{
H.~Baum, T.~Friedrich, R.~Grunewald, I.~Kath,
``Twistors and Killing Spinors on Riemannian Manifolds,''
Teubner (1991).
}

\lref\doprg{
T.~Friedrich,
``Dirac Operators in Riemannian Geometry,''
American Mathematical Society (2000).
}

\lref\MoroianuUN{
  A.~Moroianu,
  ``Parallel and Killing spinors on spin(c) manifolds,''
Commun.\ Math.\ Phys.\  {\bf 187}, 417 (1997)..
}

\lref\WittenEV{
  E.~Witten,
  ``Supersymmetric Yang-Mills theory on a four manifold,''
J.\ Math.\ Phys.\  {\bf 35}, 5101 (1994).
[hep-th/9403195].
}

\lref\SamtlebenGY{
  H.~Samtleben and D.~Tsimpis,
  ``Rigid supersymmetric theories in 4d Riemannian space,''
[arXiv:1203.3420 [hep-th]].
}

\lref\JiaHW{
  B.~Jia and E.~Sharpe,
  ``Rigidly Supersymmetric Gauge Theories on Curved Superspace,''
[arXiv:1109.5421 [hep-th]].
}

\lref\KallenNY{
  J.~Kallen,
  ``Cohomological localization of Chern-Simons theory,''
JHEP {\bf 1108}, 008 (2011).
[arXiv:1104.5353 [hep-th]].
}

\lref\RomelsbergerEG{
  C.~Romelsberger,
  ``Counting chiral primaries in N = 1, d=4 superconformal field theories,''
Nucl.\ Phys.\ B {\bf 747}, 329 (2006).
[hep-th/0510060].
}

\lref\Klare{
 C.~Klare, A.~Tomasiello and A.~Zaffaroni,
  ``Supersymmetry on Curved Spaces and Holography,''
[arXiv:1205.1062 [hep-th]].
}

\lref\WittenXJ{
  E.~Witten,
  ``Topological Sigma Models,''
Commun.\ Math.\ Phys.\  {\bf 118}, 411 (1988)..
}

\lref\WittenZZ{
  E.~Witten,
  ``Mirror manifolds and topological field theory,''
In *Yau, S.T. (ed.): Mirror symmetry I* 121-160.
[hep-th/9112056].
}

\lref\PestunRZ{
  V.~Pestun,
  ``Localization of gauge theory on a four-sphere and supersymmetric Wilson loops,''
[arXiv:0712.2824 [hep-th]].
}

\lref\KapustinKZ{
  A.~Kapustin, B.~Willett and I.~Yaakov,
  ``Exact Results for Wilson Loops in Superconformal Chern-Simons Theories with Matter,''
JHEP {\bf 1003}, 089 (2010).
[arXiv:0909.4559 [hep-th]].
}

\lref\JafferisUN{
  D.~L.~Jafferis,
  ``The Exact Superconformal R-Symmetry Extremizes Z,''
[arXiv:1012.3210 [hep-th]].
}

\lref\HamaAV{
  N.~Hama, K.~Hosomichi and S.~Lee,
  ``Notes on SUSY Gauge Theories on Three-Sphere,''
JHEP {\bf 1103}, 127 (2011).
[arXiv:1012.3512 [hep-th]].
}

\lref\DolanRP{
  F.~A.~H.~Dolan, V.~P.~Spiridonov and G.~S.~Vartanov,
  ``From 4d superconformal indices to 3d partition functions,''
Phys.\ Lett.\ B {\bf 704}, 234 (2011).
[arXiv:1104.1787 [hep-th]].
}

\lref\GaddeIA{
  A.~Gadde and W.~Yan,
  ``Reducing the 4d Index to the $S^3$ Partition Function,''
[arXiv:1104.2592 [hep-th]].
}

\lref\ImamuraUW{
  Y.~Imamura,
  ``Relation between the 4d superconformal index and the $S^3$ partition function,''
JHEP {\bf 1109}, 133 (2011).
[arXiv:1104.4482 [hep-th]].
}

\lref\ImamuraWG{
  Y.~Imamura and D.~Yokoyama,
  ``N=2 supersymmetric theories on squashed three-sphere,''
Phys.\ Rev.\ D {\bf 85}, 025015 (2012).
[arXiv:1109.4734 [hep-th]].
}

\lref\MartelliFU{
  D.~Martelli, A.~Passias and J.~Sparks,
  ``The Gravity dual of supersymmetric gauge theories on a squashed three-sphere,''
[arXiv:1110.6400 [hep-th]].
}

\lref\MartelliFW{
  D.~Martelli and J.~Sparks,
  ``The Nuts and Bolts of Supersymmetric Gauge Theories on Biaxially squashed Three-Spheres,''
[arXiv:1111.6930 [hep-th]].
}

\lref\KarlhedeAX{
  A.~Karlhede and M.~Rocek,
  ``Topological Quantum Field Theory And N=2 Conformal Supergravity,''
Phys.\ Lett.\ B {\bf 212}, 51 (1988)..
}

\lref\LawsonYR{
  H.~B.~Lawson and M.~L.~Michelsohn,
  ``Spin geometry,''
(Princeton mathematical series. 38).
}

\lref\LuNU{
  H.~Lu, C.~N.~Pope and J.~Rahmfeld,
  ``A Construction of Killing spinors on S**n,''
J.\ Math.\ Phys.\  {\bf 40}, 4518 (1999).
[hep-th/9805151].
}

\lref\BlauXG{
  M.~Blau,
  ``Killing spinors and SYM on curved spaces,''
JHEP {\bf 0011}, 023 (2000).
[hep-th/0005098].
}

\lref\WessCP{
  J.~Wess, J.~Bagger,
  ``Supersymmetry and Supergravity,''
Princeton, Univ. Pr. (1992).
}

\lref\SenPH{
  D.~Sen,
  ``Supersymmetry In The Space-time R X S**3,''
Nucl.\ Phys.\ B {\bf 284}, 201 (1987)..
}

\lref\StephaniTM{
  H.~Stephani, D.~Kramer, M.~A.~H.~MacCallum, C.~Hoenselaers and E.~Herlt,
  ``Exact solutions of Einstein's field equations,''
Cambridge, UK: Univ. Pr. (2003) 701 P.
}

\lref\OhtaEV{
  K.~Ohta and Y.~Yoshida,
  ``Non-Abelian Localization for Supersymmetric Yang-Mills-Chern-Simons Theories on Seifert Manifold,''
[arXiv:1205.0046 [hep-th]].
}



\rightline{PUPT-2414}
\Title{
} {\vbox{\centerline{Exploring Curved Superspace}}}

\centerline{ Thomas T. Dumitrescu,$^{1}$ Guido Festuccia,$^2$ and Nathan Seiberg\hskip1pt $^{2}$}
\bigskip
 \centerline{$^{1}$ {\it Department of Physics, Princeton University, Princeton, NJ 08544, USA}}
  \centerline{$^{2}${\it
Institute for Advanced Study, Princeton, NJ 08540, USA}}

\vskip50pt

\noindent We systematically analyze Riemannian manifolds~$\CM$ that admit rigid supersymmetry, focusing on four-dimensional~$\CN=1$ theories with a~$U(1)_R$ symmetry. We find that~$\CM$ admits a single supercharge, if and only if it is a Hermitian manifold. The supercharge transforms as a scalar on~$\CM$. We then consider the restrictions imposed by the presence of additional supercharges. Two supercharges of opposite~$R$-charge exist on certain fibrations of a two-torus over a Riemann surface. Upon dimensional reduction, these give rise to an interesting class of supersymmetric geometries in three dimensions. We further show that compact manifolds admitting two supercharges of equal~$R$-charge must be hyperhermitian. Finally, four supercharges imply that~$\CM$ is locally isometric to~$\CM_3 \times \R$, where~$\CM_3$ is a maximally symmetric space.

\vskip35pt

\Date{May 2012}

\newsec{Introduction}

In this paper, we present a systematic analysis of Riemannian manifolds that admit rigid supersymmetry, focusing on four-dimensional~$\CN=1$ theories with a~$U(1)_R$ symmetry. We can place any such theory on a Riemannian manifold~$\CM$ by minimally coupling it to the metric. The resulting theory is invariant under supersymmetry variations with spinor parameter~$\zeta$,\foot{The spinor~$\zeta$ is left-handed and carries un-dotted indices, $\zeta_\alpha$. Right-handed spinors are distinguished by a tilde and carry dotted indices, $\t \zeta^\alphadot$. Our conventions are summarized in appendix~A.} as  long as~$\zeta$ is covariantly constant,
\eqn\csp{\grad_\mu \zeta = 0~.}
The presence of a covariantly constant spinor dramatically restricts the geometry of~$\CM$, and it is not necessary in order to preserve supersymmetry. In many cases it is possible to place the theory on~$\CM$ in a certain non-minimal way, such that it is invariant under some appropriately modified supersymmetry variations. In this case the differential equation satisfied by the spinor~$\zeta$ is a generalization of~\csp.

Several such generalizations have been considered in the literature. For instance, we can twist by a line bundle~$L$. Given a connection~$A_\mu$ on~$L$, this leads to
\eqn\tcsp{\left(\grad_\mu - i A_\mu\right) \zeta = 0~.}
This equation admits a solution if and only if~$\CM$ is K\"ahler~\LawsonYR; see also~\MoroianuUN. The relation between twisting and rigid supersymmetry on K\"ahler manifolds is discussed in~\WittenEV. A different generalization of~\csp\ arises if we set only the spin-$3 \over 2$ component of~$\grad_\mu \zeta$ to zero,
\eqn\teq{\grad_\mu \zeta = \sigma_{\mu} \t \eta~.}
The spinor~$\t \eta$ is not independent. Rather, it captures the spin-$\half$ component of~$\grad_\mu \zeta$,
\eqn\esol{\t \eta = -{1 \over 4} \t \sigma^{\mu} \grad_\mu \zeta~.}
Equation~\teq\ is known as the twistor equation. It has been studied extensively in the mathematical literature; see for instance~\refs{\kilspman,\doprg} and references therein.  Finally, we can consider the twistor equation~\teq\ in conjunction with the twist by~$L$,
\eqn\tteq{\left(\grad_\mu - i A_\mu \right) \zeta = \sigma_{\mu} \t \eta~.}
This equation clearly includes~\csp, \tcsp, and~\teq\ as special cases. It was recently studied in the context of conformal supergravity~\Klare.

As we will see below, a systematic approach to supersymmetric field theory on curved manifolds leads to a different generalization of~\csp\ and~\tcsp,
\eqn\firstsusycon{\left( \grad_{\mu} - iA_\mu\right)\zeta = - i V_{\mu}\zeta- i V^{\nu} \sigma_{\mu\nu} \zeta~.}
Here~$V^\mu$ is a smooth, conserved vector field,~$\grad_\mu V^\mu = 0$. This equation is closely related to~\tteq, although there are important differences. We can express~\firstsusycon\ as
\eqn\shifA{\big( \grad_{\mu} - i \hat A_\mu\big)\zeta = - i  \sigma_{\mu}\big(V^{\nu} \t \sigma_\nu\zeta\big)~,}
where~$\hat A_\mu =A_{\mu}-{3\over 2} V_\mu$. Therefore, every solution~$\zeta$ of~\firstsusycon\ is a solution of~\tteq. However, given a solution~$\zeta$ of~\tteq, we see from~\shifA\ that it satisfies~\firstsusycon\ only if~$\t \eta$ in~\tteq\ can be expressed in terms of a smooth conserved~$V^\mu$,
\eqn\condeq{\t \eta = V^{\nu}\t \sigma_\nu\zeta~, \qquad \grad_\mu V^\mu = 0~.}
This is always possible in a neighborhood where~$\zeta$ does not vanish. By counting degrees of freedom, we see that~$V^\mu$ is determined up to two functions, which must satisfy a differential constraint to ensure~$\grad_\mu V^\mu = 0$. Locally, any solution of~\tteq\ is therefore a solution of~\firstsusycon, as long as~$\zeta$ does not vanish. This is no longer true if~$\zeta$ has zeros, since we cannot satisfy~\condeq\ for any smooth~$V^\mu$. It is known that~\tteq\ admits nontrivial solutions with zeros; see for instance~\refs{\kilspman,\doprg}. By contrast, it is easy to show that every nontrivial solution of~\firstsusycon\ is nowhere vanishing.

We will now explain how~\firstsusycon\ arises in the study of supersymmetric field theories on Riemannian manifolds. Following~\refs{\PestunRZ\KapustinKZ\JafferisUN-\HamaAV}, much work has focused on supersymmetric theories on round spheres. (See~\refs{\LuNU, \BlauXG} for some earlier work.) Recently, it was shown that rigid supersymmetry also exists on certain squashed spheres~\refs{\HamaEA\DolanRP\GaddeIA\ImamuraUW \ImamuraWG\MartelliFU-\MartelliFW}. A systematic approach to this subject was developed in~\FestucciaWS\ using background supergravity. In ordinary supergravity, the metric~$g_{\mu\nu}$ is dynamical and belongs to a supermultiplet that also includes the gravitino~$\psi_{\mu\alpha}$ and various auxiliary fields. Here, we would like to view these fields as classical backgrounds and allow arbitrary field configurations. This can be achieved by starting with supergravity and appropriately scaling the Planck mass to infinity. Rigid supersymmetry corresponds to the subalgebra of supergravity transformations that leaves a given background invariant. This procedure captures all deformations of the theory that approach the original flat-space theory at short distances. (There are known modifications of flat-space supersymmetry, but we will not discuss them here.) See appendix~B, which also contains a brief review of~\FestucciaWS.

In this paper, we will discuss~$\CN=1$ theories in four dimensions. The corresponding supergravity has several presentations, which differ in the choice of propagating and auxiliary fields.  Since we do not integrate out the auxiliary fields, these formulations are not equivalent and can lead to different backgrounds with rigid supersymmetry. We will focus on theories with a~$U(1)_R$ symmetry, which can be coupled to the new minimal formulation of supergravity~\refs{\SohniusTP,\SohniusFW}.\foot{The corresponding analysis for old minimal supergravity~\refs{\StelleYE,\FerraraEM} will be described in~\Oldmin. See also~\refs{\JiaHW,\SamtlebenGY}.} In this formulation, the auxiliary fields in the supergravity multiplet consist of an Abelian gauge field~$A_\mu$ and a two-form gauge field~$B_{\mu\nu}$.  The dual field strength~$V^\mu$ of~$B_{\mu\nu}$ is a well-defined, conserved vector field,
\eqn\bfs{V^\mu={1\over 2} \ep^{\mu\nu\rho\lambda}\d_\nu B_{\rho\lambda}~, \qquad \grad_\mu V^\mu = 0~.}
The gauge field~$A_\mu$ couples to the~$U(1)_R$ current of the field theory, which leads to invariance under local~$R$-transformations.

In new minimal supergravity, the variation of the gravitino takes the form
\eqn\gravvar{\eqalign{&\delta \psi_{\mu}=-2 \left(\grad_{\mu} - i A_\mu\right)\zeta - 2 i V_\mu \zeta -2 i V^{\nu} \sigma_{\mu\nu}\zeta~,\cr
&\delta\t \psi_{\mu}=-2 \left( \grad_{\mu} + i A_\mu\right) \t \zeta+2i V_\mu \t\zeta + 2 i V^{\nu} \t \sigma_{\mu\nu}\t  \zeta~.}}
The spinor parameters~$\zeta$ and~$\t \zeta$ have~$R$-charge~$+1$ and~$-1$ respectively. In Lorentzian signature, left-handed and right-handed spinors are exchanged by complex conjugation and the background fields~$A_\mu$ and~$V_\mu$ are real. This is not the case in Euclidean signature, where~$\zeta$ and~$\t \zeta$ are independent and the background fields~$A_\mu$ and~$V_\mu$ may be complex. However, we will always take the metric~$g_{\mu\nu}$ to be real.

A given configuration of the background fields~$g_{\mu\nu}, A_\mu$, and~$V_\mu$ on~$\CM$ preserves rigid supersymmetry, if and only if both variations in~\gravvar\ vanish for some choice of~$\zeta$ and~$\t \zeta$. Moreover, we can always consider variations of definite~$R$-charge. A supercharge~$\delta_\zeta$ of~$R$-charge~$+1$ corresponds to a solution~$\zeta$ of
\eqn\susycon{\left( \grad_{\mu} - iA_\mu\right)\zeta = - i V_{\mu}\zeta- i V^{\nu} \sigma_{\mu\nu}\zeta~,}
 while a supercharge~$\delta_{\,\, \tilde{} \!\!\zeta}$ of~$R$-charge~$-1$ corresponds to a solution~$\t \zeta$ of
\eqn\susyconii{\left( \grad_{\mu} + i A_\mu\right) \t \zeta =   iV_{\mu}\t \zeta +i V^{\nu} \t \sigma_{\mu\nu}\t  \zeta~.}
Note that the presence of rigid supersymmetry does not depend on the details of the field theory, since~\susycon\ and~\susyconii\ only involve supergravity background fields. From the algebra of local supergravity transformations~\refs{\SohniusTP,\SohniusFW}, we find that the commutation relations satisfied by the supercharges corresponding to~$\zeta$ and~$\t \zeta$ take the form
\eqn\algnm{\eqalign{& \{\delta_\zeta,\delta_{\,\, \tilde{} \!\!\zeta}\}=2 i \delta_K~,\cr
& \{\delta_\zeta,\delta_\zeta\} = \{\delta_{\,\, \tilde{} \!\!\zeta},\delta_{\,\, \tilde{} \!\!\zeta}\}=0~,\cr
& [\delta_K, \delta_\zeta] = [\delta_K, \delta_{\,\, \tilde{} \!\!\zeta}] = 0~.}}
The fact that~$\delta_\zeta^2 = 0$ follows from the~$R$-symmetry. If~$\t \zeta$ is absent, this comprises the entire superalgebra. In the presence of~$\t \zeta$, we can form a complex vector~$K = K^\mu \d_\mu$ with~$K^\mu =\zeta \sigma^\mu \t \zeta$ and~$\delta_K$ is the variation generated by the~$R$-covariant Lie derivative along~$K$. When acting on objects of~$R$-charge~$q$, it is given by
\eqn\deltak{\delta_K = \CL_K^A=\CL_K-i q K^\mu A_\mu~,}
where~$\CL_K$ is the conventional Lie derivative.\foot{The Lie derivative of~$\zeta$ along a vector~$X = X^\mu \d_\mu$ is given by
\eqn\spinlie{\CL_X \zeta=X^\mu \grad_\mu \zeta -{1\over 2} \grad_\mu X_\nu\sigma^{\mu\nu} \zeta~,}
and similarly for~$\t \zeta$. See appendix~A.} As we will see below,~$K$ is a Killing vector. The fact that~$\delta_K$ commutes with~$\delta_\zeta$ and~$\delta_{\,\, \tilde{} \!\!\zeta}$ is required for the consistency of~\algnm.\foot{If there are other supercharges, which correspond to additional solutions~$\eta$ or~$\t \eta$ of~\susycon\ or~\susyconii, the Killing vector~$K$ need not commute with them,
\eqn\comketa{[\delta_K, \delta_\eta] = -\delta_{\CL^A_K \eta}~, \qquad [\delta_K, \delta_{\char126\!\!\! \eta} ] = - \delta_{\CL^A_K {\char126\!\!\! \eta}}~.}}

In this paper, we will analyze Riemannian four-manifolds~$\CM$ that admit one or several solutions of~\susycon\ and~\susyconii. In section~2, we will discuss the various objects that appear in these equations, and comment on some of their general properties that will be used subsequently. The equations~\susycon\ and~\susyconii\ do not admit solutions for arbitrary values of~$g_{\mu\nu}, A_\mu$, and~$V_\mu$. This is due to the fact that they are partial differential equations, which are only consistent if the background fields satisfy certain integrability conditions. Additionally, there may be global obstructions. We would like to understand the restrictions imposed by the presence of one or several solutions, and formulate sufficient conditions for their existence.

In section~3, we show that~$\CM$ admits a single scalar supercharge, if and only if it is Hermitian. In this case, we can rewrite~\susycon\ as
\eqn\susyconc{\left(\grad^c_\mu - A_\mu^c\right)\zeta = 0~,}
where~$\grad_\mu^c$ is the Chern connection adapted to the complex structure and~$A_\mu^c$ is simply related to~$A_\mu$. The ability to cast~\susycon\ in this form crucially relies on the presence of~$V_\mu$, which is related to the torsion of the Chern connection. On a K\"ahler manifold~\susyconc\ reduces to~\tcsp. More generally, it allows us to adapt the twisting procedure of~\WittenEV\ to Hermitian manifolds that are not K\"ahler. As we will see, the auxiliary fields~$A_\mu$ and~$V_\mu$ are not completely determined by the geometry. This freedom, which resides in the non-minimal couplings parametrized by~$A_\mu$ and~$V_\mu$, reflects the fact that we can place a given field theory on~$\CM$ in several different ways, while preserving one supercharge (see appendix~B).

In section~4, we consider manifolds admitting two solutions~$\zeta$ and~$\t \zeta$ of opposite~$R$-charge. As was mentioned above, we can use them to construct a complex Killing vector~$K^\mu =~\!\! \zeta \sigma^\mu \t \zeta$. This situation is realized on any Hermitian manifold with metric
\eqn\metellintro{ds^2 =\Omega(z,\bar z)^2\left( (dw +h(z,\bar z) dz)(d\bar w +\b h(z,\bar z) d\bar z)+ c(z,\bar z)^2 dz d\bar z\right)~,}
where~$w, z$ are holomorphic coordinates. The metric~\metellintro\ describes a two-torus fibered over a Riemann surface~$\Sigma$ with metric~$d s^2_\Sigma = \Omega^2 c^2 dz d\b z$. As in the case of a single supercharge,~$\zeta$ and~$\t \zeta$ turn out to be scalars on~$\CM$. Upon dimensional reduction, they give rise to two supercharges on Seifert manifolds that are circle bundles over~$\Sigma$. Rigid supersymmetry on such manifolds was recently discussed in~\refs{\KallenNY,\OhtaEV}. Reducing once more, we make contact with the~$A$-twist on~$\Sigma$~\refs{\WittenXJ,\WittenZZ}.

Section~5 describes manifolds admitting two supercharges of equal~$R$-charge. This case turns out to be very restrictive. When~$\CM$ is compact, we will show that it must be hyperhermitian. Using the classification of~\Boyertwo, this allows us to constrain~$\CM$ to be one of the following: a flat torus~$T^4$, a~$K3$ surface with Ricci-flat K\"ahler metric, or~$S^3 \times S^1$ with the standard metric~$ds^2 = d\tau^2 + r^2 d\Omega_3$ and certain quotients thereof. We also comment on the non-compact case, which is less constrained.

In section~6 we describe manifolds admitting four supercharges. They are locally isometric to~$\CM_3\times \R$, where~$\CM_3$ is one of the maximally symmetric spaces~$S^3$, $T^3$, or~$H^3$. (The size of~$\CM_3$ does not vary along~$\R$.) In this case, the auxiliary fields~$A_\mu$ and~$V_\mu$ are tightly constrained.

We conclude in section~7 by considering several explicit geometries that illustrate our general analysis. Our conventions are summarized in appendix~A. In appendix~B we review the procedure of~\FestucciaWS\ to place a four-dimensional~$\CN=1$ theory on a Riemannian manifold~$\CM$ in a supersymmetric way, focusing on theories with a~$U(1)_R$ symmetry. Appendix~C contains some supplementary material related to section~4.

{\it Note: While this paper was being completed, we became aware of~\Klare, which has some overlap with our work. We are grateful to the authors for sharing their draft prior to publication.}

\newsec{General Properties of the Equations}

In this section we will lay the groundwork for our discussion of the equations~\susycon\ and~\susyconii,
\eqn\susyconrep{\eqalign{
&\left( \grad_{\mu} - iA_\mu\right)\zeta = - i V_{\mu}\zeta- i V^{\nu} \sigma_{\mu\nu}\zeta~,\cr
& \left( \grad_{\mu} + i A_\mu\right) \t \zeta =   iV_{\mu} \t \zeta+i V^{\nu} \t \sigma_{\mu\nu}  \t \zeta~.}}
We will study them on a smooth, oriented, connected four-manifold~$\CM$, endowed with a Riemannian metric~$g_{\mu\nu}$. The Levi-Civita connection is denoted by~$\grad_\mu$.  As we have explained in the introduction, the background fields~$A_\mu$ and~$V_\mu$ are generally complex, and~$V^\mu$ is conserved, $\grad_\mu V^\mu = 0$. Note that the equations~\susyconrep\ are invariant under
\eqn\eqinv{\zeta \rightarrow \zeta^\dagger~, \qquad \t \zeta \rightarrow {\t \zeta}^\dagger~, \qquad A_\mu \rightarrow - \b A_\mu~, \qquad V_\mu \rightarrow -\b V_\mu~.}

Under local frame rotations~$SU(2)_+ \times SU(2)_-$, the spinors~$\zeta$ and~$\t \zeta$ transform as~$(\half, 0)$ and~$(0, \half)$. Additionally, they carry charge~$+1$ and~$-1$ under conventional~$R$-transformations. Locally, the equations~\susyconrep\ are also invariant under complexified~$R$-transformations, and this is reflected in various formulas below. However, we will not make use of such transformations. (One reason is that they could lead to pathologies in the field theory.) Therefore, the real part of~$A_\mu$ transforms as a gauge field for the local~$U(1)_R$ symmetry, while the imaginary part is a well-defined one-form. In summary,~$\zeta$ is a section of~$L \otimes S_+$, where~$L$ is a unitary line bundle and~$S_+$ is the bundle of left-handed spinors, and~$\t \zeta$ is a section of~$L^{-1} \otimes S_-$ with~$S_-$ the bundle of right-handed spinors. The transition functions of~$L$ consist of local~$U(1)_R$ transformations, and the connection on~$L$ is given by the real part of~$A_\mu$.

Let us briefly comment on some global properties of the various objects introduced above. (For a more thorough discussion, see for instance~\LawsonYR.) If~$\CM$ is a spin manifold, we can choose well-defined bundles~$S_\pm$. In this case the line bundle~$L$ is also well defined. In general, an oriented Riemannian four manifold does not possess a spin structure. It does, however, admit a spin$^c$ structure. In this case it is possible to define well-behaved product bundles~$L \otimes S_+$ and~$L^{-1} \otimes S_-$, even though~$S_\pm$ and~$L$ do not exist. However, even powers of~$L$ are well defined.

Since the equations~\susyconrep\ are linear, the solutions have the structure of a complex vector space, which decomposes into solutions~$\zeta$ with~$R$-charge~$+1$ and solutions~$\t \zeta$ with~$R$-charge~$-1$. The fact that the equations are also first-order, with smooth coefficients, implies that any solution is determined by its value at a single point. Therefore, any nontrivial solution is nowhere vanishing, and this will be crucial below. Moreover, there are at most two solutions of~$R$-charge~$+1$, and likewise for~$R$-charge~$-1$.

The equations~\susyconrep\ do not admit solutions for arbitrary values of~$g_{\mu\nu}, A_\mu$, and~$V_\mu$. This is due to the fact that they are partial differential equations, which are only consistent if the background fields satisfy certain integrability conditions. Additionally, there may be global obstructions. Before attempting to solve the equations in general, we will analyze the restrictions on the background fields due to the presence of one or several solutions. Given one or several spinors satisfying~\susyconrep, it is useful to construct spinor bilinears, and these will feature prominently in our analysis. Here we will introduce various interesting bilinears and list some of their properties. These follow only from Fierz identities and do not make use of the equations~\susyconrep. We will only need the fact that the spinors are non-vanishing.

Given a spinor~$\zeta \in L \otimes S_+$, its norm~$|\zeta|^2$ is a scalar. More interestingly, we can define a real, self-dual two-form,
\eqn\cs{J_{\mu\nu} = {2 i \over |\zeta|^2} \zeta^\dagger \sigma_{\mu\nu} \zeta~,}
which satisfies
\eqn\acs{{J^\mu}_\nu {J^\nu}_\rho = - {\delta^\mu}_\rho~.}
Therefore,~${J^\mu}_\nu$ is an almost complex structure, which splits the complexified tangent space at every point into holomorphic and anti-holomorphic subspaces. The holomorphic tangent space has the following useful characterization~\LawsonYR: a vector~$X^\mu$ is holomorphic with respect to~${J^\mu}_\nu$ if and only if~$X^\mu \t \sigma_\mu \zeta = 0$.\foot{To see this, we can multiply~$X^\nu \t \sigma_\nu \zeta = 0$ by~$\zeta^\dagger \sigma^\mu$ and use~\cs\ to obtain~${J^\mu}_\nu X^\nu = i X^\mu$. Conversely, if~$X^\mu$ is holomorphic then~$\zeta^\dagger \sigma^\mu \t \sigma_\nu \zeta X^\nu = 0$. Multiplying by~$\b X_\mu$ we find~$|X^\mu \t \sigma_\mu \zeta|^2 = 0$, and hence~$X^\mu \t \sigma_\mu \zeta = 0$.}

We can also define another complex bilinear,
\eqn\pdef{P_{\mu\nu} = \zeta \sigma_{\mu\nu} \zeta~,}
which is a section of~$L^2 \otimes \Lambda^2_+$, where~$\Lambda^2_+$ denotes the bundle of self-dual two-forms. We find that
\eqn\holp{{J_\mu}^\rho P_{\rho\nu} = i P_{\mu\nu}~,}
and hence~$P_{\mu\nu}$ is anti-holomorphic with respect to the almost complex structure~${J^\mu}_\nu$.

Suppose we are given another spinor~$\t \zeta \in L^{-1} \otimes S_-$. Then we can define an anti-self-dual two-form,
\eqn\csbar{{\, \t J}_{\mu\nu} = {2 i \over |\t \zeta|^2} \t \zeta^\dagger \t \sigma_{\mu\nu} \t \zeta~.}
Again, we find that~${\, \, \t J^\mu}_{\nu} {\,\, \t J^\nu}_\rho = - {\delta^\mu}_\rho$, so that~${\, \t J^\mu}_{ \nu}$ is another almost complex structure. The two almost complex structures~${J^\mu}_\nu$ and~${\, \t J^\mu}_\nu$ commute,
\eqn\commjjt{{J^\mu}_\nu {\, \t J^\nu}_\rho - {\, \t J^\mu}_\nu {J^\nu}_\rho = 0~.}
Combining~$\zeta$ and~$\t \zeta$, we can also construct a complex vector~$K = K^\mu \d_\mu$ with
\eqn\kdef{K^\mu = \zeta \sigma^\mu \t \zeta~.}
It squares to zero, $K^\mu K_\mu = 0$, and it is holomorphic with respect to both~${J^\mu}_\nu$ and~${\, \t J^\mu}_{ \nu}$,
\eqn\kholc{{J^\mu}_\nu K^\nu = {\, \t J^\mu}_{ \nu} K^\nu = i K^\mu~.}
The norm of~$K$ is determined by the norms of~$\zeta$ and~$\t \zeta$,
\eqn\norms{\b K^{\mu} K_\mu = 2 |\zeta|^2 |\t \zeta|^2~.}
It will be useful to express~$J_{\mu\nu}$ and~${\, \t J}_{\mu\nu}$ directly in terms of~$K_\mu$,
\eqn\qdef{\eqalign{& J_{\mu\nu} = Q_{\mu\nu} + \half \ep_{\mu\nu\rho\lambda} Q^{\rho\lambda}~,\cr
& \, \t J_{\mu\nu} =Q_{\mu\nu} - \half \ep_{\mu\nu\rho\lambda} Q^{\rho\lambda}~,\cr
& Q_{\mu\nu} = { i \over  \b K^{\lambda} K_\lambda} \left(K_\mu \b K_\nu - K_\nu \b K_\mu\right)~.}}

Finally, we consider two spinors~$\zeta, \eta \in L \otimes S_+$. As above, they give rise to almost complex structures,
\eqn\jonei{ {J^{\mu}}_\nu = {2 i \over |\zeta|^2} \zeta^\dagger {\sigma^\mu}_\nu \zeta~, \qquad {I^\mu}_\nu = {2 i \over |\eta|^2} \eta^\dagger {\sigma^\mu}_\nu \eta~.}
Their anticommutator is given by
\eqn\JIc{\eqalign{& {J^\mu}_\nu  {I^\nu}_{\rho}+{I^\mu}_\nu  {J^\nu}_{\rho}=-2 f {\delta^{\mu}}_{\rho}~,\cr
& f = 2 \, {|\zeta^\dagger \eta|^2 \over |\zeta|^2 |\eta|^2 }-1~.}}
It follows from the Cauchy-Schwarz inequality that~$-1 \leq f \leq 1$, so that~$f =1$ if and only if~$\zeta$ is proportional to~$\eta$. In this case~${J^\mu}_\nu={I^\mu}_\nu$. Similarly, $f = -1$ if and only if~$\zeta$ is proportional to~$\eta^\dagger$, so that~${J^\mu}_\nu=-{I^\mu}_\nu$. By appropriately choosing independent solutions~$\zeta$ and~$\eta$ of~\susyconrep\ we can always arrange for~$f\neq \pm 1$ at a given point. This fact will be used in section~5.

\newsec{Manifolds Admitting One Supercharge}

In this section we will analyze manifolds~$\CM$ that admit a solution~$\zeta$ of~\susycon,
\eqn\susyconbis{\left( \grad_{\mu} - iA_\mu\right)\zeta = - i V_{\mu}\zeta- i V^{\nu} \sigma_{\mu\nu}\zeta~.}
The presence of such a solution implies that~$\CM$ is Hermitian. Conversely, we will show that a solution exists on any Hermitian manifold.

\subsec{Restrictions Imposed by~$\zeta$}

In section~2 we used the fact that solutions of~\susyconbis\ are nowhere vanishing to construct various bilinears out of $\zeta$, and we established some of their properties at a fixed point on~$\CM$. Here we will use the fact that $\zeta$ satisfies \susyconbis\ to study their derivatives. We begin by proving that the almost complex structure~${J^\mu}_\nu$ defined in~\cs\ is integrable, so that~$\CM$ is a complex manifold with Hermitian metric~$g_{\mu\nu}$. It suffices to show that the commutator of two holomorphic vector fields is also holomorphic. Recall from section~2 that a vector~$X^\mu$ is holomorphic with respect to~${J^\mu}_\nu$ if and only if~$X^\mu \t \sigma_\mu \zeta = 0$. By differentiating this formula, contracting with another holomorphic vector~$Y^\mu$, and antisymmmetrizing, one finds that~$[X,Y]$ is holomorphic if and only if~\LawsonYR
\eqn\xycom{X^{[\mu} Y^{\nu]} \t \sigma_\mu \grad_\nu \zeta = 0~.}
Using~\susyconbis\ and the fact that~$X^\mu, Y^\mu$ are holomorphic, we find that this is indeed the case, and hence~${J^\mu}_\nu$ is integrable.

Alternatively, we can use~\susyconbis\ to compute~$\grad_\mu {J^\nu}_\rho$ directly (this is straightforward but tedious), and show that the Nijenhuis tensor of~${J^\mu}_\nu$ vanishes,
\eqn\njt{{N^\mu}_{\nu\rho} = {J^\lambda}_\nu \grad_\lambda {J^\mu}_\rho - {J^\lambda}_\rho \grad_\lambda {J^\mu}_\nu -{J^\mu}_\lambda \grad_\nu {J^\lambda}_\rho + {J^\mu}_\lambda \grad_\rho {J^\lambda}_\nu = 0~.}
Again, it follows that the almost complex structure~${J^\mu}_\nu$ is integrable.

Using the complex structure, we can introduce local holomorphic coordinates~$z^i~(i~\!\!=~\!\!1,2)$. We will denote holomorphic and anti-holomorphic indices by un-barred and barred lowercase Latin letters respectively. In these coordinates, the complex structure takes the form,
\eqn\csholc{{J^i}_j = i {\delta^i}_j~, \qquad {J^{\ibar}}_{\jbar} = - i {\delta^{\ibar}}_{\jbar}~.}
Lowering both indices, we obtain the K\"ahler form of the Hermitian manifold,
\eqn\kf{J_{i \b j} = - i g_{i \b j}~.}
It is a real~$(1,1)$ form. The K\"ahler form~$J_{\mu\nu}$ is not covariantly constant with respect to the Levi-Civita connection, unless the manifold is K\"ahler. Instead, we compute using~\susyconbis,
\eqn\gradJmu{\grad_\mu {J^\mu}_\nu = - (V_\nu + \b V_\nu) + i  (V_\mu - \b V_\mu) {J^\mu}_\nu~.}
This implies that~$V_\mu$ takes the form
\eqn\gradJ{V_\mu = - \half \grad_\nu {J^\nu}_\mu + U_\mu~, \qquad {J_\mu}^\nu U_\nu = i U_\mu~.}
Since~$U_\mu$ only has anti-holomorphic components $U_{\b i}\,$, we see that~$V_{\b i}$ is not determined by~${J^\mu}_\nu$. This freedom in~$V_\mu$ was already mentioned in the introduction, where it reflected an ambiguity in passing from~\tteq\ to~\firstsusycon; see also the discussion in appendix~B. Imposing conservation of~$V_\mu$ leads to
\eqn\ucons{\grad^\mu U_\mu = 0~.}
Recall from~\bfs\ that~$V_\mu$ is the dual field strength of a two-form gauge field~$B_{\mu\nu}$. We can then express~\gradJ\ as~$B_{\mu\nu} = \half J_{\mu\nu} + \cdots~$, where the ellipsis denotes additional terms that reflect the freedom in~$V_{\b i}$.

Since~$\CM$ is Hermitian, it is natural to adopt a connection that is compatible with both the metric~$g_{\mu\nu}$ and the complex structure~${J^\mu}_\nu$. As we remarked above, this is not the case for the Levi-Civita connection~$\grad_\mu$, unless the manifold is K\"ahler. We will instead use the Chern connection~$\grad^c_\mu$, which has the property that~$\grad^c_\mu \, g_{\nu\rho} = 0$ and~$\grad^c_\mu {J^\nu}_\rho = 0$. This corresponds to replacing the ordinary spin connection~$\omega_{\mu\nu\rho}$ by
\eqn\chernc{\omega^c_{\mu\nu\rho} = \omega_{\mu\nu\rho} - \half {J_\mu}^\lambda \left(\grad_\lambda J_{\nu\rho} + \grad_\nu J_{\rho\lambda} + \grad_\rho J_{\lambda\nu}\right)~.}
Rewriting the spinor equation~\susyconbis\ in terms of the Chern connection, we obtain
\eqn\scc{\left(\grad^c_\mu - i A^c_\mu\right)\zeta = 0~,}
where we have defined
\eqn\ac{A^c_\mu = A_\mu +{1 \over 4} \left({\delta_\mu}^\nu - i {J_\mu}^\nu\right) \grad_\rho {J^\rho}_\nu - {3 \over 2} U_\mu~.}
Note that~$A^c_\mu$ and~$A_\mu$ only differ by a well-defined one-form, and hence they shift in the same way under~$R$-transformations.

To summarize, a solution~$\zeta$ of~\susyconbis\ defines an integrable complex structure~${J^\mu}_\nu$ and an associated Chern connection. In turn, the spinor~$\zeta$ is covariantly constant with respect to the Chern connection twisted by~$A^c_\mu$ in~\ac.

When~$\CM$ is K\"ahler, the Chern connection coincides with the Levi-Civita connection, and~$A^c_\mu = A_\mu$ if we choose~$U_\mu = 0$. In this case~\scc\ reduces to~\tcsp,
\eqn\tcspbis{\left(\grad_\mu - i A_\mu\right) \zeta = 0~.}
Conversely, it is well-known that this equation admits a solution on any K\"ahler manifold~\LawsonYR. Intuitively, this follows from the~$\CN=1$ twisting procedure described in~\WittenEV. On a K\"ahler manifold, the holonomy of the Levi-Civita connection is given by~$U(2) = U(1)_+ \times SU(2)_-$ with~$U(1)_+ \subset SU(2)_+$. For an appropriate choice of~$U(1)_R$ connection~$A_\mu$, we can cancel the~$U(1)_+$ component of the spin connection to obtain a scalar supercharge on~$\CM$. Similarly, it was shown in~\KarlhedeAX\ that the~$\CN=2$ twisting procedure of~\WittenZE\ can be interpreted in terms of a certain generalization of~\tcspbis.

Equation~\scc\ allows us to generalize this argument to an arbitrary Hermitian manifold. Given a complex structure~${J^\mu}_\nu$, the holonomy of the Chern connection is contained in~$U(2)$. As above, we can twist by~$A_\mu^c$ to obtain a solution~$\zeta$, which transforms as a scalar. This solution is related to the complex structure as in~\cs. Choosing~$V_\mu$ as in~\gradJ\ and~$A_\mu$ as in~\ac, we see that~$\zeta$ also satisfies~\susyconbis. Therefore, we can solve~\susyconbis\ on any Hermitian manifold to obtain a scalar supercharge. We will describe the explicit solution in the next subsection. Here we will explore some of its properties, assuming that it exists.

Consider~$P_{\mu\nu} = \zeta \sigma_{\mu\nu} \zeta$, which was defined in~\pdef. Note that~$P_{\mu\nu}$ locally determines~$\zeta$ up to a sign. It follows from~\holp\ that~$P_{\mu\nu}$ is a nowhere vanishing section of~$L^2 \otimes \b \CK$, where~$\b \CK = \Lambda^{0,2}$ is the anti-canonical bundle of~$(0,2)$ forms. This implies that the line bundle~$L^2 \otimes \b \CK$ is trivial, and hence we can identify~$L =(\b \CK)^{-{1\over 2}}$, up to a trivial line bundle. If~$\CM$ is not spin, the line bundle~$(\b \CK)^{-{1\over 2}}$ is not globally well defined. However, it does correspond to a good spin$^c$ structure on~$\CM$.

More explicitly, we work in a patch with coordinates~$z^i$ and define~$p  = P_{\b 1 \b2}$. Since the induced metric on~$\b \CK$ is given by~$1 \over \sqrt g$ with~$g=\det(g_{\mu\nu})$, it follows that~${1 \over \sqrt g} |p|^2$ is a positive scalar on~$\CM$. We are therefore led to consider
\eqn\prel{s = p \, g^{-{1 \over 4}}~,}
which is nowhere vanishing and has~$R$-charge~2. Under holomorphic coordinate changes~$s$ transforms by a phase,
\eqn\schange{z'^i = z'^i(z)~, \qquad s'(z') = s(z) \bigg(\det\bigg({\d z'^i \over \d z^j}\bigg) \bigg)^{\half}  \bigg(\det\bigg( {\b {\d z'^i \over \d z^j}} \bigg) \bigg)^{-\half}~.}
We can locally compensate these phase rotations by appropriate~$R$-transformations. Under these combined transformations~$s$ transforms as a scalar. Starting from a section~$p$ of the trivial line bundle~$L^2 \otimes \b \CK$ and dividing by a power of the trivial determinant bundle, we have thus produced a scalar~$s$. As we will see in the next subsection, the scalar~$s$ determines the scalar supercharge corresponding to~$\zeta$.

We will now solve for~$A_\mu$ in terms of~$s$. It follows from~\scc\ that
\eqn\pcc{\left(\grad^c_\mu - 2 i A_\mu^c\right) p = 0~.}
The Chern connection acts on sections of the anti-canonical bundle in a simple way,
\eqn\gradp{\grad^c_i p = \d_i p~, \qquad \grad^c_{\b i} p = \d_{\b i} p - {p \over 2} \d_{\b i} \log g~.}
Substituting into~\pcc\ and using~\prel, we obtain~$A_\mu^c$ and hence~$A_\mu$,
\eqn\Asolb{\eqalign{& A_\mu = A^c_\mu - {1 \over 4} \left({\delta_\mu}^\nu - i {J_\mu}^\nu\right) \grad_\rho {J^\rho}_\nu + {3 \over 2} U_\mu~,\cr
&A^c_i=-{i\over 8} \d_i \log g -{i\over 2} \d_i \log s~,\cr
& A^c_{\b i}= {i\over 8}\partial_{\b i} \log g -{i\over 2} \d_{\b i} \log s~.}}
Note that $s$ appears in \Asolb\ as the parameter of complexified local~$R$-transformations.

\subsec{Solving for~$\zeta$ on a Hermitian Manifold}

We will now show that it is possible to solve the equation~\susyconbis\ on a general Hermitian manifold~$\CM$, given its metric~$g_{\mu\nu}$ and complex structure~${J^\mu}_\nu$. The solution is not completely determined by these geometric structures. It also depends on a choice of conserved, anti-holomorphic~$U_\mu$ and a complex, nowhere vanishing scalar~$s$ on~$\CM$. In terms of this additional data, the background fields~$V_\mu$ and~$A_\mu$ are given by~\gradJ\ and~\Asolb.

We will work in a local frame that is adapted to the Hermitian metric on~$\CM$. This corresponds to a choice of vielbein~$e^1,e^2\in \Lambda^{(1,0)}$ and~$e^{\b1},e^{\b 2}\in \Lambda^{(0,1)}$, which satisfies
\eqn\vielbein{ds^2=e^1 e^{\b 1}+e^2 e^{\b 2}~.}
Any two such frames are related by a transformation in~$U(2) \subset SU(2)_+ \times SU(2)_-$. Since~\vielbein\ is preserved by parallel transport with the Chern connection, we see that its holonomy is contained in~$U(2)$. More explicitly, we choose
\eqn\frameh{
{1 \over \sqrt 2} e^1=\sqrt{g_{1\b1}} \, dz^1+{g_{2\b 1}\over \sqrt{g_{1\b 1}}} \, dz^2, \qquad {1 \over \sqrt 2} e^2={g^{1\over 4}\over \sqrt{g_{1\b 1}}} \, dz^2~.}
In this frame, the solution of~\susyconbis\ with our choice of background fields is given by\eqn\zetasol{\zeta_\alpha ={\sqrt s\over 2}\pmatrix{0 \cr 1}~.}
The complex structure is then indeed given by~$\zeta$ as in~\cs.

We have specified that~$s$ is a scalar on~$\CM$, yet~$\zeta$ in~\zetasol\ only depends on~$s$. We will now discuss the transformation properties of~$\zeta$, and explain to what extent it can be considered a scalar as well. Under a holomorphic coordinate change~$z'^i = z'^i(z)$, the metric and the vielbein transform in the usual way. In the~$z'$-coordinates, we can also define another frame~$f'^1, f'^2$, which is related to~$g'_{i \b j}$ as in~\frameh. In this frame, the spinor~$\zeta'$ takes the same form as in~\zetasol. The frames~$f'$ and~$e'$ are related by a matrix~$\CU \in U(2)$ via~$f' = \CU e'$. To relate the spinors~$\zeta'$ and~$\zeta$, we will only need the determinant of~$\CU$,\foot{This follows from the fact that the complex structure~${J^\mu}_\nu$ can be written in terms of~$\zeta$ as in~\cs, which implies that local~$U(2)$ frame rotations are identified with~$U(1)_+ \times SU(2)_- \subset SU(2)_+ \times SU(2)_-$.}
\eqn\detUzeta{\eqalign{& \zeta' = \sqrt{\det \CU} \, \zeta~,\cr
& \det \CU=  \bigg(\det\bigg({\d z'^i \over \d z^j}\bigg) \bigg)^{\half}  \bigg(\det\bigg( {\b {\d z'^i \over \d z^j}} \bigg) \bigg)^{-\half}~.}}
Hence,~$\zeta'$ and~$\zeta$ only differ by a phase, and this can be undone by an appropriate~$R$-transformation. Under this combined transformation~$\zeta$ transforms as a scalar, which is related to the scalar~$s$ via~\zetasol. Note that the phase of~$s$ can be removed by a globally well-defined~$R$-transformation. (If we were to allow complexified~$R$-transformations, we could set~$s = 1$ everywhere on~$\CM$.)

\subsec{Restrictions Imposed by~$\t \zeta$}

It is straightforward to repeat the analysis above in the presence of a solution~$\t \zeta$ of~\susyconii. As before, the complex structure~${\, \t J^\mu}_{ \nu}$ in~\csbar\ is integrable and determines the holomorphic part of~$V_\mu$,
\eqn\gradJb{V_\mu = {1 \over 2} \grad_\nu {\, \t J^\nu}_\mu + \t U_\mu~, \qquad {\, \t J_\mu}^\nu \t U_\nu = i \t U_\mu~, \qquad \grad^\mu\t U_\mu = 0~.}
The gauge field~$A_\mu$ then takes the form
\eqn\Asolbar{\eqalign{& A_\mu = A^c_\mu + {1 \over 4} \left({\delta_\mu}^\nu - i {\, \t J_\mu}^\nu\right) \grad_\rho {\, \t J^\rho}_\nu + {3 \over 2} \t U_\mu~,\cr
&A^c_i={i\over 8} \d_i \log g +{i\over 2} \d_i \log {\t s}~,\cr
& A^c_{\b i}= -{i\over 8}\partial_{\b i} \log g +{i\over 2} \d_{\b i} \log {\t s}~.}}
As above, $\t s$ is a complex scalar that determines~$\t \zeta$.

\newsec{Manifolds Admitting Two Supercharges of Opposite~$R$-Charge}

In this section we will consider manifolds~$\CM$ on which it is possible to find a pair~$\zeta$ and~$\t \zeta$ that solves the equations in~\susycon\ and~\susyconii,
\eqn\susycontris{\eqalign{
&\left( \grad_{\mu} - iA_\mu\right)\zeta = - i V_{\mu}\zeta- i V^{\nu}  \sigma_{\mu\nu}\zeta~,\cr
& \left( \grad_{\mu} + i A_\mu\right) \t \zeta=   iV_{\mu}\t \zeta +i V^{\nu} \t \sigma_{\mu\nu}\tilde  \zeta~.}}
Again, we begin by analyzing the restrictions imposed by the presence of~$\zeta$ and~$\t \zeta$, before establishing sufficient conditions for their existence. As discussed in the introduction, the solutions~$\zeta$ and~$\t\zeta$ give rise to a Killing vector field~$K=K^\mu\d_\mu$ with~$K^\mu=\zeta \sigma^\mu \t \zeta$. Together with its complex conjugate~$\bar K$, it generates part of the isometry group of $\CM$. There are two qualitatively different cases depending on whether~$K$ and~$\b K$ commute. In this section we will discuss the case when they do commute, and we will show that~$\CM$ can be described as a fibration of a torus~$T^2$ over an arbitrary Riemann surface~$\Sigma$. The non-commuting case turns out to be very restrictive. It is discussed in sections~6 and~7, as well as appendix~C.

\subsec{Restrictions Imposed by~$\zeta$ and~$\t \zeta$}

We begin by assuming the existence of two spinors~$\zeta$ and~$\t \zeta$ that solve the equations~\susycontris. From the analysis of the previous section we know that they give rise to two complex structures~${J^\mu}_\nu$ and~${\, \t J^\mu}_{ \nu}$, both of which are compatible with the metric. Recall from section~2 that the nowhere vanishing complex vector field~$K^\mu=\zeta \sigma^\mu \t \zeta$ is holomorphic with respect to both complex structures. We can now use the fact that~$\zeta$ and~$\t \zeta$ satisfy the equations~\susycontris\ to show that~$K$ is a Killing vector,
\eqn\propKprel{\eqalign{&\grad_\mu K_\nu+\grad_\nu K_\mu=0~.}}

The fact that~$K^\mu K_\mu=0$ allows us to constrain the algebra satisfied by~$K$ and its complex conjugate~$\b K$ (see appendix~C). When they do not commute, there are additional Killing vectors and the equations~\susycontris\ imply that the manifold is locally isometric to~$S^3\times \R$ with metric
\eqn\metcylpo{ds^2=d \tau^2 + r^2 d\Omega_3~.}
Here~$d\Omega_3$ is the round metric on the unit three-sphere. This case will be discussed in sections 6 and 7.

In the remainder of this section we will analyze the case in which the Killing vector~$K$ commutes with its complex conjugate $\b K$,
\eqn\propK{\eqalign{& \b K^{\nu}\grad_\nu K^\mu -K^{\nu}\grad_\nu \b K^{ \mu}=0~.}}

Using the complex structure~${J^\mu}_\nu$, we can introduce holomorphic coordinates~$w, z$. Since~$K$ is holomorphic with respect to~${J^\mu}_\nu$ and satisfies~\propK, we can choose these coordinates so that~$K  = \d_w$. The metric then takes the form
\eqn\metell{ds^2 =\Omega(z,\bar z)^2\left( (dw +h(z,\bar z) dz)(d\bar w +\b h(z,\bar z) d\bar z)+ c(z,\bar z)^2 dz d\bar z\right)~.}
The conformal factor~$\Omega^2$ is determined by the norm of~$K$, which in turn depends on the norms of~$\zeta$ and~$\t \zeta$ as in~\norms,
\eqn\omkn{\Omega^2 = 2 \b K^{\mu} K_\mu = 4 |\zeta|^2 |\b\zeta|^2~.}
The metric~\metell\ describes a two-torus~$T^2$ fibered over a Riemann surface~$\Sigma$ with metric~$d s^2_\Sigma = \Omega^2 c^2 dz d\b z$. As we will see below, the metric~\metell\ admits a second compatible complex structure, which can be identified with~${\, \t J^\mu}_{ \nu}$.

We will now constrain the form of the background field~$V_\mu$. First note that $\grad_\nu {J^\nu}_\mu~=~- \grad_\nu {\, \t J^\nu}_{ \mu}$, which follows from~\propKprel,~\propK, and the expressions~\qdef\ for~${J^\mu}_\nu$ and~${\, \t J^\mu}_{ \nu}$ in terms of~$K$. Since $\zeta$ is a solution of \susycontris, it must be that $V_\mu$ satisfies \gradJ. Similarly \gradJb\ must hold because $\tilde \zeta$ is also a solution. Consistency of these expressions requires the two conserved vectors $U_\mu$ and $\tilde U_\mu$ to satisfy
\eqn\consV{U_\mu=\tilde U_\mu~,\quad {J_\mu}^\nu U_\nu=i U_\mu~,\quad {\,\t J_\mu}^\nu \tilde U_\nu=i \tilde U_\mu~.}
Recall from~\commjjt\ that the two complex structures~${J^\mu}_\nu$ and~${\,\t J^\mu}_\nu$ commute. Moreover, they have opposite self-duality, so that the space of vectors that are holomorphic under both is one dimensional. Hence, $U_\mu=\t U_\mu$ must be proportional to~$K_\mu$ everywhere. In summary,
\eqn\Venc{V_\mu=-{1\over 2} \grad_\nu {J^\nu}_\mu+\kappa K_\mu = \half \grad_\nu {\, \t J^\nu}_{ \mu} + \kappa K_\mu~, \qquad K^\mu \d_\mu \kappa = 0~.}
Here~$\kappa$ is a complex scalar function on~$\CM$, which is constrained by the conservation of~$V_\mu$.

Given the form of~$V_\mu$ in~\Venc\ and the spinors~$\zeta$ and~$\t \zeta$, the gauge field~$A_\mu$ is completely determined. It is given by~\Asolb, or alternatively~\Asolbar. It can be checked that the consistency of these two equations does not impose additional restrictions on the metric or the background fields. This also follows from the explicit solution presented in the next subsection.

\subsec{Solving for~$\zeta$ and~$\t \zeta$}

Here we will establish a converse to the results of the previous subsection: we can find a pair~$\zeta$ and~$\t \zeta$ that solves the equations~\susycontris\ whenever the metric~$g_{\mu\nu}$ admits a complex Killing vector~$K$ that squares to zero, $K^\mu K_\mu = 0$, and commutes with its complex conjugate as in~\propK. Note that we do not assume that~$\CM$ is Hermitian. Instead, we can use~$K$ to define~${J^\mu}_\nu$ and~${\, \t J^\mu}_{ \nu}$ through the formula~\qdef, without making reference to~$\zeta$ and~$\t \zeta$. Since~$K^\mu K_\mu = 0$ these are indeed almost complex structures, and~$K$ is holomorphic with respect to both. Using~\propKprel\ and~\propK, we can show that they are integrable, i.e.\ their Nijenhuis tensor~\njt\ vanishes. Choosing complex coordinates adapted to~${J^\mu}_\nu$, the metric takes the same form as in~\metell,
\eqn\metellbis{ds^2 =\Omega(z,\bar z)^2\left( (dw +h(z,\bar z) dz)(d\bar w +\b h(z,\bar z) d\bar z)+ c(z,\bar z)^2 dz d\bar z\right)~.}

In order to exhibit the explicit solution for~$\zeta$ and~$\tilde \zeta$, we introduce a local frame adapted to the Hermitian metric~\metellbis\ as in~\frameh,
\eqn\vieher{e^1=\Omega(dw+ h dz)~, \qquad e^2=\Omega c dz~.}
Choosing the background fields~$V_\mu$ and~$A_\mu$ as in~\Venc\ and~\Asolb, we solve for~$\zeta$ and~$\t \zeta$,
\eqn\spinsol{\zeta_{\alpha}={\sqrt s \over 2} \pmatrix{0 \cr 1}~, \qquad \t \zeta^{\alphadot}= {\Omega \over \sqrt s} \pmatrix{0 \cr 1}~.}
As before, we have the freedom of choosing a nowhere vanishing complex~$s$, which transforms as a scalar under holomorphic coordinate changes followed by appropriate~$R$-transformations. Hence,~$\zeta$ can be regarded as a scalar, and the same is true for~$\t \zeta$, since~$\Omega$ is a scalar as well. (Recall from~\omkn\ that it is proportional to the norm of~$K$.) The freedom in choosing~$s$ reflects the underlying invariance of the equations~\susycontris\ under complexified~$R$-transformations, and as above we could use this freedom to set~$s = 1$.

We would like to comment on the isometries generated by~$K$ and~$\b K$. Recall that~$K$ appeared on the right-hand-side of the supersymmetry algebra~\algnm. This is not the case for~$\b K$. Nevertheless, both~$K$ and~$\b K$ are Killing vectors, because the metric is real. However,~$\b K$ need not be a symmetry of the auxiliary fields~$V_\mu$ and~$A_\mu$. For instance, to ensure that~$V_\mu$ in~\Venc\ commutes with~$\b K$ we must impose an additional restriction on~$\kappa$,
\eqn\extracon{\b K^{\mu} \d_\mu \kappa = 0~.}
Similarly, to ensure that~$A_\mu$ is invariant under~$K$ and~$\b K$ up to ordinary gauge transformations, we must impose
\eqn\extraconi{K^\mu \d_\mu |s| = \b K^{\mu} \d_\mu |s| = 0~.}
Note that~$A_\mu$ is always invariant under~$K$ up to complexified gauge transformations. The conditions~\extracon\ and~\extraconi\ ensure that~$K$ and~$\b K$ are good symmetries of all background fields. Although this choice is natural, we are free to consider auxiliary fields that are not invariant under~$\b K$.

If we choose to impose~\extracon\ and~\extraconi, we would like to add~$\b K$ to the supersymmetry algebra~\algnm. When acting on objects of~$R$-charge~$q$, we define
\eqn\deltakb{\delta_{\b K} = \CL^{\b A}_{\b K} = ~\CL_{\b K} - i q \b K^\mu \b A_\mu~,}
which is similar to~\deltak, except that we use~$\b A_\mu$ instead of~$A_\mu$. This is covariant under ordinary gauge transformations. With this definition, we find that
\eqn\kbcom{\eqalign{& [\delta_{\b K}, \delta_\zeta] = [\delta_{\b K}, \delta_{\,\,{\tilde{}}\!\! \zeta}] = 0~,\cr
& [\delta_K, \delta_{\b K} ] = 0~.}}
Together with~\algnm, these commutation relations comprise the familiar two-dimensional~$(2,0)$ supersymmetry algebra. Here it acts on the~$T^2$ fibers in~\metell.

\subsec{Trivial Fibrations and Dimensional Reduction}

Here we will comment on the case when one or both cycles of the torus are trivially fibered over the base~$\Sigma$. By reducing along these cycles, we obtain manifolds admitting two supercharges in three and two dimensions.

Let us consider the case when one of the cycles is trivially fibered, so that the manifold is of the form~$\CM_3 \times S^1$. The three-manifold~$\CM_3$ is itself a circle bundle over the Riemann surface~$\Sigma$. Let us choose~$K = \d_\tau + i \d_\psi$, where the real coordinates~$\tau$ and~$\psi$ parametrize the trivial~$S^1$ and the circle fiber of~$\CM_3$ respectively. In this case, the metric takes the form
\eqn\met{\eqalign{& ds^2 = \Omega^2(z, \b z) d\tau^2 + ds^2_{\CM_3}~,\cr
& ds^2_{\CM_3} = \Omega^2(z, \b z) \left( \big(d \psi  + a(z, \b z) dz + \b a(z,\b z) d\b z\big)^2 + c^2(z, \b z) dz d \b z\right)~.}}
Since~$K = \d_\tau + i \d_\psi$ squares to zero and commutes with its complex conjugate, this metric is in the class considered in the previous subsection. Hence, we can find two solutions~$\zeta$ and~$\t \zeta$. Imposing the additional conditions~\extracon\ and~\extraconi\ ensures that the spinors and the background fields do not vary along the two circles parametrized by~$\tau$ and~$\psi$.

We can now reduce along~$\tau$ to obtain two scalar supercharges on Seifert manifolds that are circle bundles over a Riemann surface~$\Sigma$, as long as the metric is invariant under translations along the fiber. Rigid supersymmetry on such manifolds was recently discussed in~\refs{\KallenNY,\OhtaEV}. The supercharges we find exist in any three-dimensional~$\CN=2$ theory with a~$U(1)_R$ symmetry.

If we choose the circle bundle to be trivial, we can reduce once more and obtain two scalar supercharges on any Riemann surface~$\Sigma$. They are analogous to the ones obtained by the~$A$-twist of a two-dimensional~$\CN=(2,2)$ theory on~$\Sigma$~\refs{\WittenXJ,\WittenZZ}.

\newsec{Manifolds Admitting Two Supercharges of Equal~$R$-Charge}

In this section we analyze manifolds that admit two independent solutions of~\susycon. The presence of these two solutions turns out to be very restrictive. If~$\CM$ is compact, we will prove that it must be one of the following:
\medskip
\item{$\bullet$} A torus~$T^4$ with flat metric.
\item{$\bullet$} A~$K3$ surface with Ricci-flat K\"ahler metric.
\item{$\bullet$} Certain discrete quotients of~$S^3\times S^1$ with the standard metric~$ds^2 = d\tau^2 + r^2 d\Omega_3$.
\medskip

Given two independent solutions~$\zeta$ and~$\eta$ of~\susycon, we derive a set of consistency conditions for the metric and the auxiliary fields. Using~${1\over 2} R_{\mu\nu\kappa\lambda} \sigma^{\kappa \lambda}\zeta = [\grad_\mu,\grad_\nu]\zeta$ and the fact that~$\zeta$ satisfies~\susycon, we obtain
\eqn\consa{\eqalign{{1\over 2} R_{\mu\nu\kappa\lambda} \sigma^{\kappa \lambda}\zeta =~& V^\rho V_\rho \sigma_{\mu\nu}\zeta +i (\d_\mu (A_\nu-V_\nu)-\d_\nu(A_\mu-V_\mu))\zeta\cr
&-i (\grad_{\mu} +i V_\mu)V^{\rho} \sigma_{\nu \rho} \zeta+i (\grad_{\nu} +i V_\nu)V^{\rho} \sigma_{\mu \rho} \zeta~,}}
and similarly for~$\eta$. Since~$\zeta$ and~$\eta$ are independent at every point, we arrive at the following integrability conditions:
\medskip
\item{1.)} The Weyl tensor is anti-self-dual, $W_{\mu\nu\rho\lambda}  =  - \half \ep_{\mu\nu\kappa\sigma} {W^{\kappa\sigma}}_{\rho\lambda}$.

\item{2.)} The curl of~$V_\mu$ is anti-self-dual, $\d_\mu V_\nu-\d_\nu V_\mu= -\half \ep_{\mu\nu\rho\lambda} (\d^\rho V^\lambda - \d^\lambda V^\rho)$.

\item{3.)} The difference $A_\mu-V_\mu$ is closed, $\d_\mu (A_\nu - V_\nu) - \d_\nu (A_\mu - V_\mu) = 0$.

\item{4.)} The Ricci tensor is given by
\eqn\Riccnd{R_{\mu\nu}=i (\grad_{\mu}V_{\nu} + \grad_\nu V_\mu)-{2} (V_{\mu}V_\nu-g_{\mu\nu} V_\rho V^\rho)~.}
\medskip
\noindent If we instead consider two independent solutions~$\t \zeta$ and~$\t \eta$ of~\susyconii, the Weyl tensor and the curl of~$V_\mu$ in~$1.)$ and~$2.)$ are self-dual rather than anti-self-dual, and the Ricci tensor is given by~\Riccnd\ with~$V_\mu \rightarrow -V_\mu$.

These conditions locally constrain the geometry of the manifold. They take a particularly simple form on manifolds of~$SU(2)$ holonomy, which are Ricci-flat and have anti-self-dual Weyl tensor. In this case, we can satisfy the integrability conditions by choosing~$V_\mu = A_\mu = 0$. Indeed, such manifolds admit two independent covariantly constant spinors. Further examples are discussed in sections~6 and~7.

Here we will not attempt to classify all manifolds that satisfy the conditions above. Instead, we will focus on the case when~$\CM$ is compact, and prove the following global result: the existence of two spinors~$\zeta$ and~$\eta$ that satisfy~\susycon\ everywhere on a compact manifold~$\CM$ implies that~$\CM$ is hyperhermitian. Compact hyperhermitian four-manifolds have been classified in~\Boyertwo. Up to a global conformal transformation, they are given by the manifolds listed at the beginning of this section.\foot{See also the discussion in~\IvanovAI, where these manifolds are identified with compact Hermitian surfaces for which the restricted holonomy of the Bismut connection is contained in~$SU(2)$.} Using the fact that~$V_\mu$ is conserved, we will find that the conformal factor must be a constant.

A hyperhermitian structure on $\CM$ arises whenever there are two anti-commuting hermitian structures ${J^{(1)}}$ and ${J^{(2)}}$. Together with their commutator $J^{(3)}$ they satisfy the quaternion algebra,
\eqn\quatalg{\{ J^{(a)}, J^{(b)}\} = -2 \delta^{ab}~, \qquad (a, b = 1, 2, 3)~.} This implies that there is an entire $S^2$ of Hermitian structures parametrized by
\eqn\cssphere{J(\vec n)=\sum_{a} n^a J^{(a)}~,\qquad |\vec n|=1~.}

Since~$\zeta$ and~$\eta$ satisfy~\susycon, the almost complex structures~${J^\mu}_\nu$ and ${I^\mu}_\nu$ constructed in~\jonei\ are integrable. Recall from~\JIc\ that the anticommutator of~${J^\mu}_\nu$ and~${I^\mu}_\nu$ gives rise to a real function~$f$, which is determined in terms of~$\zeta$ and~$\eta$. Moreover, we are free to choose~$\zeta$ and~$\eta$ such that~$f\neq\pm 1$ at a given point; at this point ${J^\mu}_\nu \neq \pm {I^\mu}_\nu$. We will now prove that ${J^\mu}_\nu$ and ${I^\mu}_\nu$ are elements of a hyperhermitian structure~on~$\CM$.

In order to establish this result, we will consider the Lee forms associated with~${J^\mu}_\nu$ and~${I^\mu}_\nu$,
\eqn\lfdef{\eqalign{&\theta^{J}_\mu={J_\mu}^\rho \grad_\nu {J^\nu}_\rho~,\cr
& \theta^I_\mu={I_\mu}^\rho \grad_\nu {I^\nu}_\rho~.}}
Using \gradJmu\ they can be expressed as follows:
\eqn\Leeforms{\eqalign{ &\theta^{J}_\mu=i(V_\mu-\b V_\mu)-{J_\mu}^\rho(V_\rho+\b V_\rho)~,\cr
&\theta^I_\mu=i(V_\mu-\b V_\mu)-{I_\mu}^\rho(V_\rho+\b V_\rho)~.}}
We will also need  the following formula, which follows from~\susycon\ by direct computation:
\eqn\normdiff{\d_\mu \log {|\zeta|^2 \over |\eta|^2} = - \half \left(J_{\mu\nu} - I_{\mu\nu}\right) \big(V^\nu + \b V^{\nu}\big)~.}
Subtracting the two equations in~\Leeforms\ and using~\normdiff, we find that the Lee forms differ by an exact one-form,
\eqn\difftheta{\theta^J_\mu-\theta^I_\mu= \d_\mu h~, \qquad h = 2 \log {|\zeta|^2\over |\eta|^2}~.}

Recall from the first integrability condition listed above that the Weyl tensor must be anti-self-dual. We thus have a compact four-manifold~$\CM$ with anti-self-dual Weyl tensor that admits two Hermitian structures $J$ and $I$ such that $J\neq \pm I$ somewhere on $\CM$. Applying proposition~(3.7) in~\Pontecorvo, it follows that
\medskip
\item{a)} The function~$h$ in~\difftheta\ is constant.
\item{b)} The manifold admits a hyperhermitian structure as in~\quatalg. Moreover, $J$ and~$I$ belong to it, and hence they can be expressed in terms of the $J^{(a)}$ as in~\cssphere.
\medskip
\noindent We conclude that~$\CM$ is one of the manifolds listed at the beginning of this section, up to a global conformal rescaling of the metric. In order to fix the conformal factor, we use the fact that~$V_\mu$ is conserved. Observe
that since ${J^\mu}_\nu$ and ${I^\mu}_\nu$  are independent elements of the hyperhermitian structure, ${J^\mu}_\nu- {I^\mu}_\nu$ is invertible. It therefore follows from~\normdiff\ and~\difftheta\ with $\d_\mu h=0$ that $V_\mu$ is purely imaginary. Hence, we see from~\Leeforms\ that
\eqn\Leehyper{\theta^J_\mu=\theta^I_\mu =  i(V_\mu-\b V_\mu)~.}
Since $V_\mu$ is conserved,~$\theta_\mu^I$ and~$\theta_\mu^J$ are conserved as well. To see that this fixes the conformal factor, consider a conformal rescaling of the metric, $\hat g_{\mu\nu} = e^\phi g_{\mu\nu}$. It follows from \lfdef\ that the Lee forms shift by an exact one-form,
\eqn\leeshift{\hat \theta_\mu^J = \theta_\mu^J + \d_\mu \phi~,}
and similarly for~$\theta_\mu^I$. Since~$\theta_\mu^J$ is conserved, $\hat \theta_\mu^J$ can only be conserved if~$\phi$ is harmonic. On a compact manifold, this is only possible for constant~$\phi$. It can be checked that the manifolds listed at the beginning of this section all have conserved Lee forms, and hence they are the correct hyperhermitian representatives within each conformal class. This is trivial for a flat~$T^4$ or a~$K3$ surface with Ricci-flat metric, since both are~hyperk\"ahler manifolds and hence the Lee forms~\lfdef\ vanish. Even though~$S^3 \times S^1$ is not K\"ahler, it can be checked that the Lee forms are conserved if we choose the standard metric~$ds^2 = d\tau^2 + r^2 d \Omega_3$. This case will be discussed in section~7.

\newsec{Manifolds Admitting Four Supercharges}

In this section we will formulate necessary conditions for the existence of four supercharges. These follow straightforwardly from the integrability conditions discussed in section~5. Assuming the existence of two independent solutions of~\susycon, we found that the Weyl tensor and the curl of~$V_\mu$ must be anti-self-dual. Similarly, two solutions of~\susyconii\ imply that they are also self-dual, and hence they must vanish. It follows that~$\CM$ is locally conformally flat and that~$V_\mu$ is closed. Since~$A_\mu - V_\mu$ is also closed, it follows that the gauge field~$A_\mu$ is flat. Finally, the Ricci tensor must satisfy~\Riccnd\ and the same relation with~$V_\mu \rightarrow -V_\mu$. This implies that~$\grad_\mu V_\nu + \grad_\nu V_\mu = 0$, and since~$V_\mu$ is also closed, it must be covariantly constant,
\eqn\covc{\grad_\mu V_\nu=0~.}
The Ricci tensor is then given by
\eqn\Riccicontr{R_{\mu\nu}= -2 (V_{\mu}V_\nu-g_{\mu\nu} V_\rho V^\rho)~.}

Since~$V_\mu$ is covariantly constant, $\CM$ is locally isometric to~$\CM_3 \times \R$. It follows from~\Riccicontr\ that~$\CM_3$ is a space of constant curvature. Let~$r$ be a positive constant. There are three possible cases:
\medskip
\item{1.)} If~$V^\mu V_\mu=-{1\over r^2}$ then~$\CM_3$ is locally isometric to a round~$S^3$ of radius~$r$. In this case~$V^\mu$ is purely imaginary and points along~$\R$.

\item{2.)} If~$V_\mu = 0$ then~$\CM$ is locally isometric to flat~$\R^4$. This is the case of ordinary~$\CN=1$ supersymmetry in flat space.

\item{3.)} If~$V^\mu V_\mu={1\over r^2}$ then~$\CM_3$ is locally isometric to~$H^3$, the three-dimensional hyperbolic space of radius $r$ and constant negative curvature. In this case~$V^\mu$ is real and points along~$\R$.
\medskip
\noindent We will discuss cases~$1.)$ and~$3.)$ below.

\newsec{Examples}

\subsec{$S^3\times \R$}

Consider~$S^3\times \R$ with metric
\eqn\metrnd{ds^2= d\tau^2 + r^2 d\Omega_3~,}
where~$d\Omega_3$ is the round metric on a unit three sphere. As we saw in section~6, this manifold admits four supercharges. Supersymmetric field theories on this space have been studied in~\refs{\SenPH,\RomelsbergerEG} and more recently in~\FestucciaWS. Here we will examine this example from the point of view of the preceding discussion.

Since this manifold admits rigid supersymmetry, it must be Hermitian. We can introduce holomorphic coordinates~$w,z$, so that the metric~\metrnd\ takes the form
\eqn\mcyls{ds^2=\Big(dw- {r \b z \over r^2+|z|^2} \, dz\Big) \Big(d\b w- {r z\over r^2+|z|^2} \, d\b z\Big)  +{r^4 \over (r^2 +|z|^2)^2} \, dz d\b z~.}
Here the imaginary part of~$w$ is periodic, $w \sim w +2 \pi i r$.\foot{The point~$z=\infty$ is covered by different coordinates~$w', z'$,
\eqn\zzp{z'={r^2\over  z}~, \qquad w'= w -r\log { z \over r}~,}
as long as~$z \neq 0$. In these coordinates, the metric takes the same form as in~\mcyls. Due to the periodicity of~$w$, we do not need to choose a specific branch for the logarithm in~\zzp.} The vector~$\d_w+\d_{\b w}$ is covariantly constant and points along~$\R$, while~$i(\d_w-\d_{\b w})$ generates translations along a Hopf fiber of~$S^3$. Since the metric~\mcyls\ is of the form~\metell, it allows for two supercharges~$\zeta$ and~$\tilde \zeta$ such that~$K= \zeta \sigma^\mu \t \zeta \d_\mu$ is equal to the holomorphic Killing vector~$\d_w$. With the choice of frame in~\vieher, these two solutions are given by \spinsol.

As discussed in section 6, it is possible to choose the auxiliary fields~$A_\mu$ and~$V_\mu$ to obtain four supercharges on~$S^3 \times \R$. Up to a sign, this fixes
\eqn\Veqfsc{V=- {i\over r}(\d_w+\d_{\b w})~.}
We will comment on the other choice of sign below. The gauge field~$A_\mu$ must be flat, but it is otherwise undetermined, and hence we can add an arbitrary complex Wilson line for~$A_\mu$ along~$\R$. In general, the resulting supercharges vary along~$\R$. This is not the case if we choose~$A_\mu = V_\mu$, so that we can compactify to~$S^3 \times S^1$. In the frame~\vieher\ the supercharges take the form
\eqn\solsphere{\zeta_\alpha=\pmatrix{\,\,  a_1 e^{-(w-\b w)/2r} \cr \!\! a_2 e^{(w-\b w)/2r}}~,\quad \t \zeta^\alphadot=\pmatrix{ \!\! a_3 e^{( w-\b w)/2r} \cr \,\, a_4 e^{-(w- \b w)/2r}}~,}
where the~$a_i$ are arbitrary complex constants. Setting~$a_1 = a_3 = 0$, we obtain the two supercharges~$\zeta$ and~$\t \zeta$ discussed above, which are of the form~\spinsol. Since~$A_\mu$ and~$V_\mu$ are purely imaginary, we can use~\eqinv\ to obtain two other supercharges~$\zeta^\dagger$ and~$\t\zeta^\dagger$. They correspond to setting~$a_2 = a_4 = 0$ in~\solsphere.

Setting~$a_3 = a_4 = 0$, we obtain two supercharges of equal~$R$-charge on the compact manifold~$S^3 \times S^1$. This manifold is hyperhermitian but not K\"ahler. If~$V$ is given by~\Veqfsc, the Lee forms in~\Leehyper\ are non-vanishing but conserved, in accord with the general discussion in section~5.

Using the spinors~$\zeta$ and~$\t \zeta$ in~\solsphere, we can construct four independent complex Killing vectors of the form~$K^\mu = \zeta \sigma^\mu \t \zeta$. Since the supercharges are related by~$\zeta \leftrightarrow \zeta^\dagger$ and~$\t \zeta \leftrightarrow \t \zeta^\dagger$, these vectors are linear combinations of four real, orthogonal Killing vectors~$L_a~(a = 1, 2, 3)$ and~$T$, which satisfy the algebra
\eqn\commLL{[L_a,L_b]= \epsilon_{abc} L_c~, \qquad [L_a, T] = 0~.}
The~$L_a$ generate the~$SU(2)_l$ inside the~$SU(2)_l \times SU(2)_r$ isometry group of~$S^3$, while~$T$ generates translations along~$\R$. The supercharges form two~$SU(2)_l$ doublets that carry opposite~$R$-charge and are invariant under~$SU(2)_r$. (If we choose the opposite sign for~$V$ in~\Veqfsc, the spinors are invariant under~$SU(2)_l$ and transform as doublets under~$SU(2)_r$.) Using~\algnm\ and~\comketa\ we find that the supersymmetry algebra is~$SU(2|1)$.

If we remain on~$S^3 \times S^1$ but only require two supercharges~$\zeta$ and~$\t \zeta$, the auxiliary fields~$V_\mu$ and~$A_\mu$ are are less constrained. For instance, we can choose~$\zeta$ and~$\t \zeta$ corresponding to~$a_1 = a_3 = 0$ in~\solsphere. As discussed above, they give rise to the holomorphic Killing vector~$K = \d_w$, which commutes with its complex conjugate~$\b K$. According to the discussion in section 4, we can preserve~$\zeta$ and~$\t \zeta$ for any choice
\eqn\shiftaux{V= {i\over r}(\d_w+\d_{\b w})+ \kappa K~,\qquad A= {i\over r}(\d_w+\d_{\b w}) + {3\over 2} \kappa K~,}
where~$\kappa$ is a complex function that satisfies~$K^\mu\d_\mu \kappa=0$, so that~$V^\mu$ is conserved. If we instead choose~$\zeta$ and~$\t \zeta$ corresponding to~$a_1=a_4=0$ in~\solsphere, the resulting Killing vector~$K$ points along the~$S^3$. Together with its complex conjugate~$\b K$, it generates the~$SU(2)_l$ isometry subgroup. For generic choices of~$a_i$ we find that~$K$ and~$\b K$ generate the~$SU(2)_l\times U(1)$ isometry subgroup that also includes translations along~$S^1$. These possibilities for the algebra of Killing vectors are precisely the ones discussed in appendix C.

\subsec{$H^3\times \R$}

Consider~$H^3\times \R$, where~$H^3$ is the three-dimensional hyperbolic space of constant negative curvature. This manifold also admits four supercharges, and in some respects it is similar to the previous example~$S^3 \times \R$, but there are also qualitative differences.

As before, we can introduce holomorphic coordinates~$w, z$ and write the metric in the form,
\eqn\mcylh{ds^2=e^{(z+\b z)/ r} \, dw d\b w + dz d\b z~,}
where~$r$ is the radius of~$H^3$. Now the coordinates~$w, z$ cover the entire space. The covariantly constant vector~$i (\d_{z}-\d_{\b z})$ points along~$\R$. In order to preserve four supercharges, we must choose
\eqn\solVV{V= {i\over r}(\d_{z}-\d_{\b z})~,}
up to a sign (see below). As in the previous example, we are free to add an arbitrary complex Wilson line for~$A_\mu$ along~$\R$. Setting~$A_\mu=V_\mu$ and choosing the frame~\vieher, the four supercharges are given by
\eqn\solhtcyl{\zeta_\alpha=\pmatrix{a_1 e^{-(z+\b z)/4r} \cr (a_2-a_1 {w\over r}) \, e^{z+\b z/4r}}~,\qquad \t \zeta^\alphadot=\pmatrix{ a_3 e^{-(z+\b z)/4r} \cr (a_4-a_3 {w\over r})\, e^{(z+\b z)/ 4r}}~.}
As before, the~$a_i$ are arbitrary complex constants. Since the metric~\mcylh\ is of the form~\metell, two of the supercharges are given by~\spinsol. They correspond to~$a_1=a_3=0$ in~\solhtcyl. Note that the spinors in~\solhtcyl\ do not depend on~$z-\b z$, so that we can compactify to~$H^3 \times S^1$.

In contrast to~$S^3 \times S^1$, the supercharges~$\zeta^\dagger$ and~$\t \zeta^\dagger$ correspond to choosing the opposite sign in~\solVV. Now the four independent complex Killing vectors~$K^\mu = \zeta \sigma^\mu \t \zeta$ constructed from~$\zeta$ and~$\t \zeta$ in~\solhtcyl\ give rise to seven independent real Killing vectors, which comprise the full~$SL(2,\C)\times U(1)$ isometry group of~$H^3 \times S^1$. The four complex Killing vectors~$K$ commute with all four complex conjugates~$\b K$.

\subsec{Squashed~$S^3 \times \R$}

We now consider~$S^3 \times \R$, where the~$S^3$ is one of the squashed three-spheres discussed in~\HamaEA. They are defined by their isometric embedding in flat~$\R^4$, where they satisfy the constraint
\eqn\embd{ {1 \over a^2} \big(x_1^2+x_2^2\big)+{1 \over b^2} \big(x_3^2+x_4^2\big)=1~.}
Rotations in the~$x_1x_2$ and~$x_3x_4$ planes generate a~$U(1)\times U(1)$ isometry. Introducing angular coordinates~$\theta \in [0,\pi/2]$, $\alpha\sim \alpha+2\pi,$ and~$\beta\sim \beta+2\pi$ for the squashed~$S^3$ and a coordinate~$\tau$ along~$\R$, we can write the metric in the form
\eqn\squau{\eqalign{&ds^2=d\tau^2+F(\theta)^2 d\theta^2 + a^2\cos^2\!\theta d\alpha^2+b^2 \sin^2\!\theta d\beta^2~,\cr & F(\theta)=\sqrt{a^2 \sin^2\!\theta +b^2 \cos^2\!\theta}~. }}
In these coordinates, the~$U(1) \times U(1)$ isometry of the squashed sphere is generated by the real Killing vectors~$\d_\alpha$ and~$\d_\beta$. By combining them with translations~$\d_\tau$ along~$\R$, we obtain a complex Killing vector~$K$, which squares to zero and commutes with its complex conjugate,
\eqn\killvtwo{\eqalign{& K= \d_\tau- {i\over a} \partial_\alpha- {i\over b} \partial_\beta~.}}

According to the discussion in section~4, this guarantees the existence of two supercharges~$\zeta$ and~$\t \zeta$. Introducing the frame
\eqn\framesq{e^1=d\tau~, \quad e^2=F(\theta) \, d\theta~, \quad e^3=a \cos\theta\, d\alpha~, \quad e^4=b \sin\theta\, d\beta~,}
they are given by
\eqn\solone{\zeta_\alpha=-{i\over \sqrt{2}}\pmatrix{e^{{i\over 2}(\alpha+\beta-\theta)}\cr  i e^{{i\over 2}(\alpha+\beta+\theta)}}~,\qquad \t \zeta^\alphadot=-{i\over \sqrt{2}}\pmatrix{ e^{-{i\over 2}(\alpha+\beta-\theta)}\cr  i e^{-{i\over 2}(\alpha+\beta+\theta)}}~.}
The auxiliary fields take the form
\eqn\auxchoice{\eqalign{&V_\mu dx^\mu=-{i\over F(\theta)} \, d\tau+ \kappa K_\mu dx^\mu~,\qquad K^\mu \partial_\mu \kappa=0~,\cr
& A_\mu dx^\mu= -{1\over 2 F(\theta)}\, \big(2i d\tau+a d\alpha +b d\beta\big)+{1\over 2}(d\alpha+d\beta)+{3\over 2} \kappa K_\mu dx^\mu~.}}
As in section~4, we can use~$K$ to define a complex structure compatible with the metric~\squau. In the corresponding holomorphic coordinates, the metric is of the form~\metell\ and the supercharges are given by~\spinsol.

We can again fix the metric~\squau\ and obtain two supercharges for different choices of the auxiliary fields~$A_\mu$ and~$V_\mu$. For instance, we can obtain a second solution by replacing~$\tau \rightarrow -\tau$ and~$\beta \rightarrow \pi - \beta$ in~\killvtwo\ and repeating the previous construction. Note that this does not change the orientation of the manifold. From these two solutions, we can obtain two more by using~\eqinv.

\vskip 1cm

\noindent {\bf Acknowledgments:}
We thank M.~Rocek for collaboration in the early stages of this project. We also thank A.~Dymarsky and E.~Witten for useful discussions, and D.~Freed, Z.~Komargodski and M.~Rocek for comments on the manuscript.  We are especially grateful to the authors of~\Klare\ for sharing their draft prior to publication. The work of TD was supported in part by a DOE Fellowship in High Energy Theory and a Centennial Fellowship from Princeton University. The work of GF was supported in part by NSF grant PHY-0969448. TD and GF would like to thank the Weizmann Institute of Science for its kind hospitality during the completion of this project. The work of NS was supported in part by DOE grant DE-FG02-90ER40542.
Any opinions, findings, and conclusions or recommendations expressed in this
material are those of the authors and do not necessarily reflect the views of the funding agencies.

\appendix{A}{Conventions}

We follow the conventions of~\WessCP, adapted to Euclidean signature. This leads to some differences in notation, which are summarized here, together with various relevant formulas.

\subsec{Flat Euclidean Space}

The metric is given by~$\delta_{\mu\nu}$, where~$\mu, \nu = 1, \ldots, 4$. The totally antisymmetric Levi-Civita symbol is normalized so that~$\ep_{1234} = 1$. The rotation group is given by~$SO(4) = SU(2)_+ \times SU(2)_-$. A left-handed spinor~$\zeta$ is an~$SU(2)_+$ doublet and carries un-dotted indices, $\zeta_\alpha$. Right-handed spinors~$\t \zeta$ are doublets under~$SU(2)_-$. They are distinguished by a tilde and carry dotted indices, $\t \zeta^\alphadot$. In Euclidean signature, $SU(2)_+$ and~$SU(2)_-$ are not related by complex conjugation, and hence~$\zeta$ and~$\tilde \zeta$ are independent spinors.

The Hermitian conjugate spinors~$\zeta^\dagger$ and~${\t \zeta}^{\dagger}$ transform as doublets under~$SU(2)_+$ and~$SU(2)_-$ respectively. They are defined with the following index structure,
\eqn\daggers{(\zeta^\dagger)^\alpha = \overline {(\zeta_\alpha)}~, \qquad (\t \zeta^\dagger)_\alphadot = \overline {(\t \zeta^\alphadot)}~,}
where the bars denote complex conjugation. Changing the index placement on both sides of these equations leads to a relative minus sign,
\eqn\daggersii{(\zeta^\dagger)_\alpha = - \overline{(\zeta^\alpha)}~, \qquad (\t \zeta^\dagger)^\alphadot = - \overline{(\t \zeta_\alphadot)}~.}
We can therefore write the~$SU(2)_+$ invariant inner product of~$\zeta$ and~$\eta$ as~$\zeta^\dagger \eta$. Similarly, the~$SU(2)_-$ invariant inner product of~$\t \zeta$ and~$\t \eta$ is given by~${\t \zeta}^\dagger \t \eta$. The corresponding norms are denoted by~$|\zeta|^2 = \zeta^\dagger \zeta$ and~$|\t \zeta|^2 = \t \zeta^\dagger \t \zeta$.

The sigma matrices take the form
\eqn\sigmamat{\sigma^\mu_{\alpha\alphadot} = (\vector{\sigma}, -i)~,\qquad \t \sigma^{\mu \alphadot\alpha} = (-\vector{\sigma}, -i)~,}
where~$\vector{\sigma} = (\sigma^1, \sigma^2, \sigma^3)$ are the Pauli matrices. We use a tilde (rather than a bar) to emphasize that~$\sigma^\mu$ and~$\t \sigma^\mu$ are not related by complex conjugation in Euclidean signature. The sigma matrices~\sigmamat\ satisfy the identities
\eqn\antic{\sigma_\mu\t \sigma_\nu + \sigma_\nu \t \sigma_\mu = -2\delta_{\mu\nu}~, \qquad \t \sigma_\mu \sigma_\nu + \t \sigma_\nu \sigma_\mu = -2\delta_{\mu\nu}~.}
The generators of~$SU(2)_+$ and~$SU(2)_-$ are given by the antisymmetric matrices
\eqn\smunu{\sigma_{\mu\nu} = {1 \over 4} (\sigma_\mu \t \sigma_\nu - \sigma_\nu\t \sigma_\mu)~, \qquad \t \sigma_{\mu\nu} = {1 \over 4} (\t \sigma_\mu \sigma_\nu - \t \sigma_\nu \sigma_\mu)~.}
They are self-dual and anti-self-dual respectively,
\eqn\asd{\half \ep_{\mu\nu\rho\lambda} \sigma^{\rho \lambda } = \sigma_{\mu\nu}~, \qquad \half \ep_{\mu\nu\rho\lambda} \t \sigma^{\rho\lambda} = - \t \sigma_{\mu\nu}~.}

\subsec{Differential Geometry}

We will use lowercase Greek letters~$\mu, \nu, \ldots$ to denote curved indices and lowercase Latin letters~$a, b, \ldots$ to denote frame indices. Given a Riemannian metric~$g_{\mu\nu}$, we can define an orthonormal tetrad~${e^a}_\mu$. The Levi-Civita connection is denoted~$\grad_\mu$ and the corresponding spin connection is given by
\eqn\spincon{{\omega_{\mu a}}^b = {e^b}_\nu \grad_\mu {e_a}^\nu~.}
The Riemann tensor takes the form
\eqn\riemann{{R_{\mu\nu a}}^b = \d_\mu {\omega_{\nu a}}^b - \d_\nu {\omega_{\mu a}}^b + {\omega_{\nu a}}^c {\omega_{\mu c}}^b - {\omega_{\mu a}}^c {\omega_{\nu c}}^b~.}
The Ricci tensor is defined by~$R_{\mu\nu} = {R_{\mu\rho\nu}}^\rho$, and~$R = {R_\mu}^\mu$ is the Ricci scalar. Note that in these conventions, the Ricci scalar is negative on a round sphere.

The covariant derivatives of the spinors~$\zeta$ and~$\t \zeta$ are given by
\eqn\covsp{\grad_\mu \zeta = \d_\mu \zeta + \half \omega_{\mu a b } \sigma^{ab} \zeta~, \qquad \grad_\mu \t \zeta = \d_\mu \t \zeta + \half \omega_{\mu a b } {\t \sigma}^{ab} \t \zeta~.}
We will also need the commutator of two covariant derivatives,
\eqn\comspd{[\grad_\mu, \grad_\nu] \zeta = \half R_{\mu\nu a b} \sigma^{ab} \zeta~, \qquad [\grad_\mu, \grad_\nu] \t\zeta = \half R_{\mu\nu a b} {\t \sigma}^{ab} \t \zeta~.}
Finally, the Lie derivatives of~$\zeta$ and~$\t \zeta$ along a vector field~$X = X^\mu \d_\mu$ are given by~\refs{\Kosmann},
\eqn\lsp{\eqalign{& \CL_X \zeta=X^\mu \grad_\mu \zeta -{1\over 2} \grad_\mu X_\nu\sigma^{\mu\nu} \zeta~,\cr
&  \CL_X \t \zeta=X^\mu \grad_\mu \t \zeta -{1\over 2} \grad_\mu X_\nu{\t \sigma}^{\mu\nu} \t \zeta~.}}

\appendix{B}{Review of Curved Superspace}

In this appendix we explain how to place a four-dimensional~$\CN=1$ theory on a Riemannian manifold~$\CM$ in a supersymmetric way.  We review the procedure of~\FestucciaWS\ and comment on several points that play an important role in our analysis.

\subsec{Supercurrents}

Given a flat-space field theory, we can place it on~$\CM$ by coupling its energy-momentum tensor~$T_{\mu\nu}$ to the background metric~$g_{\mu\nu}$ on~$\CM$. In a supersymmetric theory, the energy-momentum tensor resides in a supercurrent multiplet~$\CS_\mu$, which also contains the supersymmetry current~$S_{\mu\alpha}$ and various other operators. In the spirit of~\SeibergVC\ we can promote the background metric~$g_{\mu\nu}$ to a background supergravity multiplet, which also contains the gravitino~$\psi_{\mu\alpha}$ and several auxiliary fields. They couple to the operators in~$\CS_\mu$.

In general, the supercurrent multiplet~$\CS_\mu$ contains~$16+16$ independent operators~\refs{\KomargodskiRB,\DumitrescuIU}. In many theories, it can be reduced to a smaller multiplet, which only contains~$12+12$ operators. There are two such~$12+12$ supercurrents: the Ferrara-Zumino~(FZ) multiplet~\FerraraPZ\ and the~$\CR$-multiplet (see for instance~\GatesNR). The~$\CR$-multiplet exists whenever the field theory possesses a~$U(1)_R$ symmetry, and this is the case we will focus on here.

The~$\CR$-multiplet satisfies the defining relations\foot{The supercovariant derivatives~$D_\alpha$ and~$\t D_\alphadot$ are given by
\eqn\scov{D_\alpha = {\d \over \d \theta^\alpha} + i \sigma^\mu_{\alpha\alphadot} \t \theta^\alphadot \d_\mu~, \qquad \t D_\alphadot = - {\d \over \d \t \theta^\alphadot} - i \theta^\alpha \sigma^\mu_{\alpha\alphadot} \d_\mu~.}}
\eqn\rmultss{\t D^\alphadot \CR_{\alpha\alphadot} = \chi_\alpha~, \qquad \t D_\alphadot \chi_\alpha = 0~, \qquad D^\alpha \chi_\alpha = \t D_\alphadot \t \chi^\alphadot~.}
Here~$\CR_{\alpha\alphadot} = - 2 \sigma^\mu_{\alpha\alphadot} \CR_\mu$ is the bi-spinor corresponding to~$\CR_\mu$. In components,
\eqn\rmultcomp{\eqalign{ \CR_\mu =~& j^{(R)}_\mu - i \theta S_\mu + i \t \theta \t S_\mu + \theta \sigma^\nu \t \theta \, \big(2 T_{\mu\nu} + {i \over 2} \ep_{\mu\nu\rho\lambda} \CF^{\rho\lambda} - {i \over 2} \ep_{\mu\nu\rho\lambda} \d^\rho j^{(R)\lambda}\big) \cr
&- \half \thetasq \t \theta \, \t \sigma^\nu \d_\nu S_\mu + \half {\t \theta\, }^2 \theta \sigma^\nu \d_\nu \t S_\mu - {1 \over 4} \thetasq {\t \theta\, }^2 \d^2 j^{(R)}_\mu~,\cr
\chi_\alpha =~& - 2i (\sigma^\mu {\t S}_\mu)_\alpha -4 \theta_\beta \left({\delta_\alpha}^\beta \, {T_\mu}^\mu - i {\left(\sigma^{\mu\nu}\right)_\alpha}^\beta \CF_{\mu\nu}\right) -4 \thetasq (\sigma^{\mu\nu} \d_\mu S_\nu)_\alpha + \cdots~.}}
Here~$j_\mu^{(R)}$ is~$R$-current, $S_{\mu\alpha}$ is the supersymmetry current, ~$T_{\mu\nu}$ is the energy-momentum tensor, and~$\CF_{\mu\nu}$ is a closed two-form, which gives rise to a string current~$\ep_{\mu\nu\rho\lambda} \CF^{\rho\lambda}$. All of these currents are conserved. Note that~\rmultcomp\ contains several unfamiliar factors of~$i$, because we are working in Euclidean signature. In Lorentzian signature, the superfield~$\CR_\mu$ is real.

It is convenient to express the closed two-form~$\CF_{\mu\nu}$ in terms of a one-form~$\CA_\mu$,
\eqn\fisda{\CF_{\mu\nu} = \d_\mu \CA_\nu - \d_\nu \CA_\mu~.}
In general~$\CA_\mu$ is not well defined, because it can shift by an exact one-form,~$\CA_\mu  \rightarrow \CA_\mu + \d_\mu \alpha$. An exception occurs if the theory is superconformal, in which case~$\CA_\mu$ is a well-defined conserved current. The superfield~$\chi_\alpha$ can then be set to zero by an improvement transformation. Below, we will need the variation of the bosonic fields in the~$\CR$-multiplet under ordinary flat-space supersymmetry transformations,
\eqn\rmctrans{\eqalign{& \delta j^{(R)}_\mu = - i \zeta S_\mu + i \t \zeta \t S_\mu~,\cr
& \delta T_{\mu\nu} = \half \zeta \sigma_{\mu\rho} \d^\rho S_\nu + \half \t \zeta \, \t\sigma_{\mu\rho} \d^\rho \t S_\nu + \left(\mu \leftrightarrow \nu\right)~,\cr
& \delta \CA_\mu = -{i\over2} \left(\zeta S_\mu - \t \zeta \t S_\mu - 2 \zeta \sigma_{\mu\rho} S^\rho + 2 \t \zeta \, \t \sigma_{\mu\rho} \t S^\rho\right) + \d_\mu \left(\cdots\right)~.}}
The ellipsis denotes a possible ambiguity in the variation of~$\CA_\mu$ due to shifts by an exact one-form.

\subsec{Background Supergravity and the Rigid Limit}

We would like to place a supersymmetric flat-space theory on a curved manifold~$\CM$ by coupling it to background supergravity fields. A straightforward but tedious approach is to follow the Noether procedure. This can be avoided if an off-shell formulation of dynamical supergravity is available. As explained in~\FestucciaWS, we can couple this supergravity to the field theory of interest and freeze the supergravity fields in arbitrary background configurations by rescaling them appropriately and sending the Planck mass to infinity. This was termed the rigid limit in~\FestucciaWS. In this limit, the fluctuations of the supergravity fields decouple and they become classical backgrounds, which can be chosen arbitrarily. In particular, we do not eliminate the auxiliary fields via their equations of motion.

We will apply this procedure to~$\CN=1$ theories in four dimensions, which admit different supercurrent multiplets. These give rise to different off-shell formulations of supergravity, which differ in the choice of propagating and auxiliary fields. For instance, the FZ-multiplet couples to the old minimal formulation of supergravity~\refs{\StelleYE,\FerraraEM}, while the~$\CR$-multiplet couples to new minimal supergravity~\refs{\SohniusTP,\SohniusFW}. We will focus on the latter. In addition to the metric~$g_{\mu\nu}$ and the gravitino~$\psi_{\mu\alpha}$, new minimal supergravity contains two auxiliary fields: an Abelian gauge field~$A_\mu$ and a two-form gauge field~$B_{\mu\nu}$. The dual field strength~$V^\mu$ of~$B_{\mu\nu}$ is a conserved vector field,
\eqn\vdapp{V^\mu={1\over 2} \ep^{\mu\nu\rho\lambda}\d_\nu B_{\rho\lambda}~, \qquad \grad_\mu V^\mu = 0~.}

In an expansion around flat space, $g_{\mu\nu} = \delta_{\mu\nu} + 2 h_{\mu\nu}$, the linearized couplings to new minimal supergravity are determined by the operators in the~$\CR$-multiplet~\rmultcomp,
\eqn\linbg{\eqalign{& {\scr L} = T_{\mu\nu} h^{\mu\nu} - j^{(R)}_\mu \big(A^\mu - {3 \over 2} V^\mu\big) - \CA_\mu V^\mu + \left({\rm fermions}\right)~.}}
Here~$A^\mu$ and~$V^\mu$ both have dimension~$1$, while~$g_{\mu\nu}$ and~$B_{\mu\nu}$ are dimensionless. The fermion terms contain the couplings of the gravitino to the supersymmetry current, which will not be important for us. We see from~\linbg\ that~$A_\mu$ is the gauge field associated with local~$U(1)_R$ transformations. Under these transformations the gravitino~$\psi_{\mu\alpha}$ has~$R$-charge~$+1$. Note that the couplings in~\linbg\ are well defined under shifts of~$\CA_\mu$ by an exact one form, because~$V^\mu$ is conserved. The fact that~$A^\mu$ and~$V^\mu$ couple to~$j_\mu^{(R)}$ and~$\CA_\mu$ is a general feature of the rigid limit that persists beyond the linearized approximation around flat space.  At higher order there are also terms quadratic in the auxiliary fields, as well as curvature terms, which are described in~\FestucciaWS.

Note that the couplings~\linbg\ do not modify the short-distance structure of the field theory, which is the same as in flat space. To see this, we can choose a point on~$\CM$ and examine the theory in Riemann normal coordinates around this point. If the curvature scale is given by~$r$, the metric is flat up to terms of order~${1 \over r^2}$. In these coordinates, the deformation~\linbg\ of the flat-space Lagrangian reduces to operators of dimension~$3$ or less, and hence the short-distance structure is not affected.

We are interested in configurations of the bosonic supergravity background fields that preserve some amount of rigid supersymmetry. The gravitino is set to zero. Such backgrounds must be invariant under a subalgebra of the underlying supergravity transformations. This subalgebra must leave the gravitino invariant,
\eqn\gravvar{\eqalign{&\delta \psi_{\mu}=-2 \left(\grad_{\mu} - i A_\mu\right)\zeta - 2 i V_\mu \zeta -2 i V^{\nu} \sigma_{\mu\nu}\zeta~,\cr
&\delta\t \psi_{\mu}=-2 \left( \grad_{\mu} + i A_\mu\right) \t \zeta+2i V_\mu \t\zeta + 2 i V^{\nu} \t \sigma_{\mu\nu}\t  \zeta~.}}
Since the variations of the bosonic supergravity fields are proportional to the gravitino, they vanish automatically. Therefore, any nontrivial choice of~$\zeta$ and~$\t \zeta$ that satisfies~\gravvar\ gives rise to a rigid supercharge. In general, the algebra satisfied by these supercharges differs from the ordinary supersymmetry algebra in flat space. Rather, it is a particular subalgebra of the local supergravity transformations that is determined by the background fields.

\subsec{Freedom in the Auxiliary Fields}

In section~3 we found that the auxiliary fields~$A_\mu$ and~$V_\mu$ are not completely determined by the geometry of the underlying Hermitian manifold. For instance, it follows from~\gradJ\ that we have the freedom of shifting~$V^\mu$ by a conserved holomorphic vector~$U^\mu$. Here we would like to elucidate the origin of this freedom by linearizing the metric around flat space, so that the deformation of the Lagrangian is given by~\linbg, and using our knowledge of the~$\CR$-multiplet.

We can choose
\eqn\onesc{\zeta_\alpha = {1 \over 2} \pmatrix{0 \cr 1}~,}
and use holomorphic coordinates~$w, z$ adapted to the complex structure defined by~$\zeta$ as in~\cs. In these coordinates, the linearized Hermitian metric only has components~$h_{i \b j}$. We would like to determine the values of the auxiliary fields~$A_\mu$ and~$V_\mu$ for which the bosonic terms in~\linbg\ are invariant under the supercharge~$\delta_\zeta$ corresponding to~$\zeta$ in~\onesc. This amounts to finding combinations of~$T_{i \b j}$ and the other bosonic operators~$j^{(R)}_\mu$,~$\CA_\mu$ in the~$\CR$-multiplet that are invariant under~$\delta_\zeta$. Since we are working to linear order, we can use the flat-space transformations~\rmctrans\ to find
\eqn\susycomb{\eqalign{& \delta_\zeta \Big(T_{w \b w} - {i \over 2} \CF_{z \b z} - {i \over 4} \big(\d_w j^{(R)}_{\b w} - \d_{\b w} j^{(R)}_w\big) + {i \over 2} \d_z j^{(R)}_{\b z}\Big) = 0~,\cr
& \delta_\zeta \Big(T_{w \b z} + {i \over 2} \CF_{w \b z} - {3 i \over 4} \d_w j^{(R)}_{\b z} + {i \over 4} \d_{\b z} j^{(R)}_w\Big) = 0~,}}
and two more with~$ w \leftrightarrow z~, \; \b w \leftrightarrow \b z$.
Moreover, up to shifts of~$\CA_\mu$ by an exact one-form,
\eqn\ainv{\delta_\zeta \CA_w = \delta_\zeta \CA_z = 0~.}
We can therefore add~$\CA_w$ and~$\CA_z$ with coefficients that are arbitrary functions of the coordinates, as long as we ensure that the Lagrangian is invariant under shifts of~$\CA_\mu$. Hence, the following Lagrangian is invariant under~$\delta_\zeta$,
\eqn\hermlinlag{\eqalign{{\scr L} =~& \Big(2  h^{w \b w} \big(T_{w \b w} - {i \over 2} \CF_{z \b z} - {i \over 4}(\d_w j^{(R)}_{\b w} - \d_{\b w} j^{(R)}_w) + {i \over 2} \d_z j^{(R)}_{\b z}\big) \cr
& + 2  h^{w\b z}  \big(T_{w \b z} + {i \over 2} \CF_{w \b z} - {3 i \over 4} \d_w j^{(R)}_{\b z} + {i \over 4} \d_{\b z} j^{(R)}_w\big)\Big) + \left(w \leftrightarrow z~, \b w \leftrightarrow \b z\right)\cr
&-U^w \CA_w -U^z \CA_z~.}}
Here~$U^{w, z}$ is a holomorphic vector field. Invariance under shifts of~$\CA_\mu$ by an exact one-form implies that it must be conserved,
\eqn\gicond{\d_w U^w + \d_z U^z = 0~.}
We can now determine the auxiliary fields~$V_\mu$ and~$A_\mu$ by comparing~\hermlinlag\ to~\linbg,
\eqn\hermlinbg{\eqalign{& V_w = -2i \left(\d_w  h_{z \b z} - \d_z  h_{w \b z}\right)~, \qquad V_{\b w} =2i \left(\d_{\b w}  h_{z \b z} - \d_{\b z}  h_{z \b w}\right) + U_{\b w}~,\cr
& A_w = -i \d_w \left( h_{w \b w} +  h_{z \b z}\right)~, \qquad A_{\b w} = i \d_{\b w} \left( h_{w \b w} +  h_{z \b z} \right) + V_{\b w}+ {1\over 2} U_{\b w}~,}}
and four more with~$w \leftrightarrow z~, \; \b w \leftrightarrow \b z$. This exactly agrees with~\gradJ\ and~\Asolb. We see that the freedom in~$U_{\b i}$ is the result of~\ainv, which allows us to add~$\CA_w$ and~$\CA_z$ with coefficients that are arbitrary functions of the coordinates, as long as we ensure invariance under shifts of~$\CA_\mu$ by an exact one-form.

\appendix{C}{Solutions with~$[K, \b K] \neq 0$}

In this appendix we analyze the constraints due to a complex Killing vector~$K$ that squares to zero, $K^\mu K_\mu=0$, and does not commute with its complex conjugate, $[K, \b K] \neq 0$. In this case, we will show that~$\CM$ is locally isometric to~$S^3\times \R$ with warped metric
\eqn\metcylcf{ds^2=d\tau^2+r(\tau)^2 d\Omega_3~.}
Here~$d\Omega_3$ is the round metric on a unit three-sphere. If~$K$ is constructed from solutions~$\zeta$ and~$\t \zeta$ of \susycontris, we further prove that~$r(\tau)$ must be a constant.

\subsec{Algebra of Killing Vectors}

For ease of notation, we will write~$\langle X, Y\rangle = X^\mu Y_{\mu}$ for any two complex vectors~$X$ and~$Y$, and refer to it as their inner product even though the vectors are complex. Since the complex conjugate~$\b K$ of the Killing vector~$K$ is also a Killing vector, their commutator gives rise to a third Killing vector $L$, which must be real,
\eqn\defJv{[K,\b K] = -i L~.}
The case~$L = 0$ was analyzed in section~4. Here we assume that~$L\neq 0$. In order to constrain the algebra generated by~$K, \b K,$ and~$L$, we will differentiate their inner products along the vectors themselves. For instance, differentiating~$\langle K,K\rangle=0$ along~$\b K$ gives
\eqn\kbkk{0 = \CL_{\b K} \langle K, K \rangle=2i \langle L ,K \rangle~.}
Since~$L$ is real, this implies that the three real Killing vectors~$K + \b K$, $i (K - \b K),$ and~$L$ are orthogonal.

We will now consider two distinct cases. If~$K, \b K,$ and~$L$ form a closed algebra, it follows from constraints similar to~\kbkk\ that this algebra must be~$SU(2)$ in its usual compact form. If the algebra does not close, we find a fourth real Killing vector, which is orthogonal to the first three. In this case the algebra is~$SU(2) \times U(1)$.

In the first case, we can introduce~$SU(2)$-invariant one-forms~$\omega^a$ and write the metric as~$ds^2 = d\tau^2 + h_{ab}(\tau) \omega^a \omega^b$. The fact that the three Killing vectors are orthogonal implies that~$h_{ab}(\tau) = r(\tau)^2 \delta_{ab}$, see for instance~\StephaniTM. The metric is therefore given by~\metcylcf\ and the isometry group is enhanced to~$SU(2) \times SU(2)$. So far, $r(\tau)$ is an arbitrary positive function.

In the second case, we can similarly show that the metric must take the form~\metcylcf, but in this case the presence of the additional~$U(1)$ isometry corresponds to translations along~$\tau$, and hence~$r(\tau)$ is a constant.

\subsec{Proof that~$r(\tau)$ is a Constant}

As we saw above, the algebra of Killing vectors is not always sufficient to prove that~$r(\tau)$ is constant. We will now show that this must be the case if~$K^\mu = \zeta \sigma^\mu \t \zeta$, where~$\zeta$ and~$\t \zeta$ are solutions of~\susycontris. If we demand that the auxiliary fields~$A_\mu$ and~$V_\mu$ respect the~$SU(2) \times SU(2)$ isometry of~\metcylcf, the equations~\susycontris\ can be analyzed explicitly and a solution exists if and only if~$r(\tau)$ is a constant. We will now give a proof that does not rely on this additional assumption.

Recall that~$\zeta$ and~$\t \zeta$ give rise to integrable complex structures~${J^\mu}_\nu$ and~${\, \t J^\mu}_\nu$, which can be expressed in terms of~$K$ and~$\b K$ as in~\qdef. Using~\gradJ\ and~\gradJb, we find that
\eqn\vuut{V_\mu = - \half \grad_\nu {J^\nu}_\mu + U_\mu = \half \grad_\nu {\, \t J^\nu}_\mu + \t U_\mu~.}
The vectors~$U^\mu$ and~$\t U^\mu$ are conserved and holomorphic with respect to~${J^\mu}_\nu$ and~${\, \t J^\mu}_\nu$ respectively. We can use the Killing vectors to parametrize them at every point:
\eqn\uut{\eqalign{& U_\mu = \kappa K_\mu + \sigma (L_\mu - i T_\mu)~,\cr
& \t U_\mu = \t \kappa K_\mu - \t \sigma (L_\mu + i T_\mu)~,}}
where~$\kappa, \sigma$ and~$\t \kappa, \t \sigma$ are complex functions on~$\CM$. Here we have defined an additional vector,\foot{It follows from the form of the metric~\metcylcf\ that~$\langle K, \b K\rangle$ is proportional to~$r^2(\tau)$. For ease of notation, we will omit an overall real constant in some of the formulas below. This will be indicated by a tilde.}
\eqn\Zvec{T={i \over  \langle K, \b K\rangle}\ep^{\mu\nu \rho \lambda} L_\nu K_\rho \b K_\lambda \d_\mu \sim  r(t) \d_\tau~.}
The fact that~$K, L - i T$ and~$K, L + i T$ are holomorphic with respect to~${J^\mu}_\nu$ and~${\, \t J^\mu}_\nu$ respectively follows from~\qdef.

Substituting~\uut\ into~\vuut\ and demanding consistency leads to
\eqn\firstcond{\sigma =\t \sigma \sim {1\over r(\tau)^2 }~,\qquad  \kappa=\t \kappa~.}
Here we have used the fact that
 \eqn\reljjb{\grad_\mu ({J^\mu}_\nu+{\, \t J^\mu}_{ \nu})= {2\over \langle K, \b K\rangle }L_\nu \sim {1\over r(\tau)^2 } L_\nu~,}
which follows from~\qdef\ and the commutation relation~\defJv. Substituting~\firstcond\ into~\uut\ and using the fact that~$U_\mu$ is conserved, we find that
\eqn\conscond{ \grad_\mu( \kappa K^\mu) \sim i {r'(\tau)\over r(\tau)^2}~.}
The orbits of~$K, \b K, L$ are given by surfaces of constant~$\tau$. Since the isometry is~$SU(2)$, they must be compact. Integrating~\conscond\ over such an orbit, we find that
\eqn\condcond{\int d^3 x \, \sqrt g \, \grad_\mu(\kappa K^\mu) \sim i r'(\tau) r(\tau) = 0~.}
Therefore~$r(\tau)$ is a constant.

\listrefs

\end